\documentclass[aps,prd,twocolumn,showpacs,superscriptaddress,nofootinbib,floatfix,showkeys,10pt]{revtex4-2}

\usepackage{graphicx}
\graphicspath{{images/}}

\usepackage{dcolumn}
\usepackage{bm}
\usepackage{amsmath,amsthm,amssymb}
\usepackage{mathrsfs}
\usepackage{braket}
\usepackage{multirow}
\usepackage{booktabs}
\usepackage{hhline}
\usepackage{makecell}
\usepackage[mathlines]{lineno}

\usepackage[colorlinks=true,linkcolor=blue,citecolor=blue,urlcolor=blue]{hyperref}

\usepackage[normalem]{ulem} 
\usepackage[dvipsnames]{xcolor} 

\makeatletter
\newcommand{\figcaption}{\def\@captype{figure}\caption}
\newcommand{\tabcaption}{\def\@captype{table}\caption}

\newcommand{\Rmnum}[1]{\expandafter\@slowromancap\romannumeral #1@}
\def\hlinewd#1{%
 \noalign{\ifnum0=`}\fi\hrule \@height #1 \futurelet
 \reserved@a\@xhline}
\makeatother

\begin{document}

\title{Isospin-breaking effects of the double-charm molecular pentaquarks}

\author{Fei-Yu Chen}
\affiliation{School of Physics, Sun Yat-Sen University, Guangzhou 510275, China
}

\author{Ning Li}
\email{lining59@mail.sysu.edu.cn}
\affiliation{School of Physics, Sun Yat-Sen University, Guangzhou 510275, China
}

\author{Wei Chen}
\email{chenwei29@mail.sysu.edu.cn}
\affiliation{School of Physics, Sun Yat-Sen University, Guangzhou 510275, China
}
\affiliation{Southern Center for Nuclear-Science Theory (SCNT), Institute of Modern Physics,
 Chinese Academy of Sciences, Huizhou 516000, Guangdong Province, China}

\begin{abstract}
 We investigate isospin-breaking effects in double-charm molecular pentaquarks with the $D^{(*)}\Sigma_c^{(*)}$ configuration, using the one-boson-exchange potential framework. In these systems, the isospin-breaking effects arise from two sources: the strong interaction, which manifests as the threshold difference of the $D^{(*)}\Sigma_c^{(*)}$
 components in the same isospin multiplet and the mass splittings of the exchanged isovector mesons ($\pi$ and $\rho$); and the electromagnetic interaction between charged $D^{(*)}$ and $\Sigma_c^{(*)}$ components. We calculate the binding properties and the isospin mixing angle between the $I=1/2$ and $I=3/2$ states of the $D^{(*)}\Sigma_c^{(*)}$ system. Our results show that the isospin-breaking effect contributes a significant correction of roughly $10\%-30\%$ to the binding energy. This effect is particularly pronounced in loosely bound molecular candidates, which are characterized by small binding energies and large root-mean-square radii. We therefore conclude that the explicit inclusion of isospin-breaking effects is essential for achieving the precision in theoretical calculations necessary to match rapidly advancing experimental programs. Our results are expected to provide valuable guidance for future high-precision experimental studies of deuteron-like molecular states.
\end{abstract}

\maketitle

\section{Introduction}\label{introduction}

The observation of the $X(3872)$ by the Belle Collaboration in
2003~\cite{Belle:2003nnu} marked the beginning of a remarkable era, triggering
a plethora of discoveries of unexpected hadron states over the following two
decades. These new hadron states challenge the conventional quark model,
including the charged charmonium-like
$Z_c(3900)$~\cite{BESIII:2013ris,Belle:2013yex}, the hidden-charm pentaquark
family $P_{\psi}(4312)$, $P_{\psi}(4440)$, $P_{\psi}(4457)$, $P_{\psi
 s}(4459)$, $P^{\Lambda}_{\psi s}(4338)$
\cite{LHCb:2015yax,LHCb:2019kea,LHCb:2020jpq,LHCb:2022ogu}, the doubly charmed
tetraquark state $T_{cc}^{+}(3875)$~\cite{LHCb:2021vvq} etc., which are
considered as candidates of exotic hadron states. Collectively, these
discoveries have reshaped our understanding of hadron spectroscopy, and for
comprehensive overviews of the experimental and theoretical progress on these
exotic states, we refer the reader to recent review
articles~\cite{Swanson:2006st, Chen_2016, Chen:2016spr, Hosaka:2016pey, Esposito:2016noz,
 Olsen:2017bmm, Brambilla:2019esw, Liu:2019zoy, HXChen_2021, Chen:2022asf, LHCb_2022}.

\begin{figure}
 \includegraphics[scale=0.50, trim={1cm 0cm 1cm 0cm}, clip]{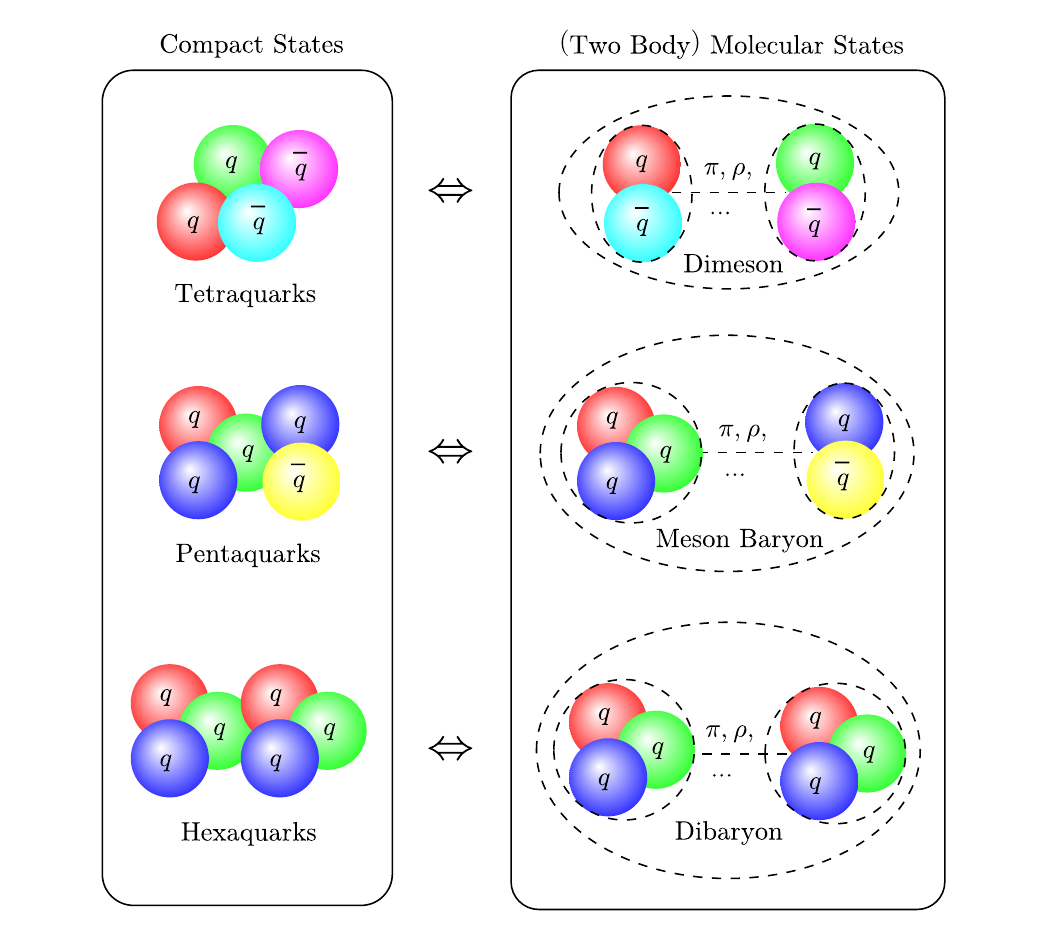}
 \caption{ \label{fig:Hadron_Molecule}
 A schematic picture of the multiquark states, compact multiquarks (left panel) versus hadronic molecules (right panel) formed by the colorless hadrons.}
\end{figure}

The experimental discoveries have prompted intensive theoretical investigations
into the nature of these states, giving rise to numerous configurational
interpretations, such as glueballs, hybrids, compact multiquarks, and hadronic
molecules. The hadronic molecule interpretation holds particular interest for
two key reasons. The first is the well-established precedent of the deuteron
itself, a proton-neutron bound state, which proves that color-singlet hadrons
can be exchanged to form molecular systems. The second is the observation that
many of the newly observed states lie remarkably close to the thresholds of two
conventional hadrons, a characteristic feature expected for loosely bound
molecular states composed of color-singlet constituents. A schematic comparison
of the compact multiquark and hadronic molecule configurations is shown in
Fig.~\ref{fig:Hadron_Molecule}.

The rapid progress in experimental studies of exotic hadrons has brought
hidden-charm tetraquarks and pentaquarks to the forefront of hadronic physics.
In fact, theoretical predictions of hidden-charm pentaquarks with the
$\bar{D}^{(*)}\Sigma_{c}^{(*)}$ molecular configuration were already made more
than a decade ago~\cite{Yang:2011wz, Wu:2010jy, Karliner:2015ina, Wang:2011rga,
 Li:2014gra, Wu:2010vk}. After the observation of the $P_{\psi}$ states in
2015~\cite{LHCb:2015yax}, numerous theoretical studies further explored their
molecular pentaquark
interpretations~\cite{Chen_2015,Chen:2015moa,Chen_2017a,Chen_2019,Chen_2016_2}.
As the doubly charmed counterpart to the hidden-charm pentaquarks, the
double-charm pentaquark $P_{cc}$ was investigated using one-boson-
exchange (OBE) model since 2010s ~\cite{Chen_2017b, Shimizu_2017, ChenRui_2021, Dong_2021}. More recently, predictions for double-charm pentaquark
states have proliferated across various theoretical frameworks, including using the QCD sum rules~\cite{Wang:2018lhz, Wang_2024, Duan_2024,Yang:2024okq,Yang:2025aer,chen2025}, the chiral effective field theory~\cite{Guo:2017vcf, ChenKan_2021, MENG20231}, the (chiral) quark model~\cite{Zhou:2018bkn, Yang_2020, Xing_2021,Li_2025, An:2025qfw}, the QDCSM method ~\cite{liu2023}, the unitarized coupled-channel approach ~\cite{Shen_2023}
and so on.

As experimental datasets continue to grow in both size and precision,
high-quality theoretical calculations are urgently needed to keep pace.
Approximations commonly employed in phenomenological studies, such as perfect
isospin symmetry, are no longer sufficient to meet the demands of future
experimental investigations. Indeed, several recent studies have highlighted
the importance of incorporating electromagnetic contributions in hadronic
systems. For instance, LQCD and OBE model analyses of the
$\Omega_{ccc}\Omega_{ccc}$ dibaryon show that while the system is bound by
several MeV through residual strong interactions alone, it becomes unbound once
electromagnetic effects are included~\cite{Lyu:2021qsh, Liu:2021pdu}. In
contrast, the $\Omega_{bbb}\Omega_{bbb}$ dibaryon remains deeply bound even
with electromagnetic interaction, exhibiting a binding energy up to $89$ MeV,
with Coulomb corrections playing only a minor role~\cite{Mathur:2022ovu}. As
for the extensively studied $X(3872)$, isospin-breaking effects modify its
binding energy by less than $1$ MeV, yet they prove essential for correctly
describing its decay patterns~\cite{Chen:2024xlw,Li_2012}. These examples
underscore the importance of accounting for isospin-breaking effects when
striving for the precision required by contemporary and forthcoming
experiments.

In the present work, we investigate isospin-breaking effects in double-charm
hadronic molecules composed of a charmed meson $D^{(*)}$ and a charmed baryon
$\Sigma_c^{(*)}$, within the framework of OBE model. To manifest the
isospin-breaking effect, we omit coupled-channel effects between distinct
flavor channels. However, we do include the mixing between different partial
waves induced by the tensor force, which couples orbital angular momentum
states differing by $\Delta l = 2$.

The paper is organized as follows. After the introduction of
Sec.~\ref{introduction}, we describe the theoretical framework including the
spin-isospin wave functions, the strong forces, coulomb forces, and the
Schr\"odinger equation with isospin breaking in
Sec.~\ref{theoretical_framework}. Our numerical results and their analyses are
presented in Sec.~\ref{numerical_results}. The last section is a brief summary
and discussion. Detailed notations and formulae used throughout the paper are
provided in the Appendix.

\section{THEORETICAL FRAMEWORK}\label{theoretical_framework}

In this section, we will present the theoretical formalism. We first give the
spin and isospin wave functions for the $D^{(*)} \Sigma_{c}^{(*)}$ systems.
Then we derive the strong interaction between a charmed meson $D^{(*)}$ and a
charmed baryon $\Sigma_c^{(*)}$, which is accounted by exchanging a light meson
(such as $\pi$, $\rho$, $\sigma$, etc.) within the one-boson-exchange potential
model. After that, we discuss the isospin breaking in detail. Lastly, we
present the matrix of the Sch\"odinger equation with isospin breaking.

\subsection{Spin and Isospin Wave Functions}

\begin{figure}[h]
 \includegraphics[scale=0.75, trim={4cm 0cm 4cm 0cm}, clip]{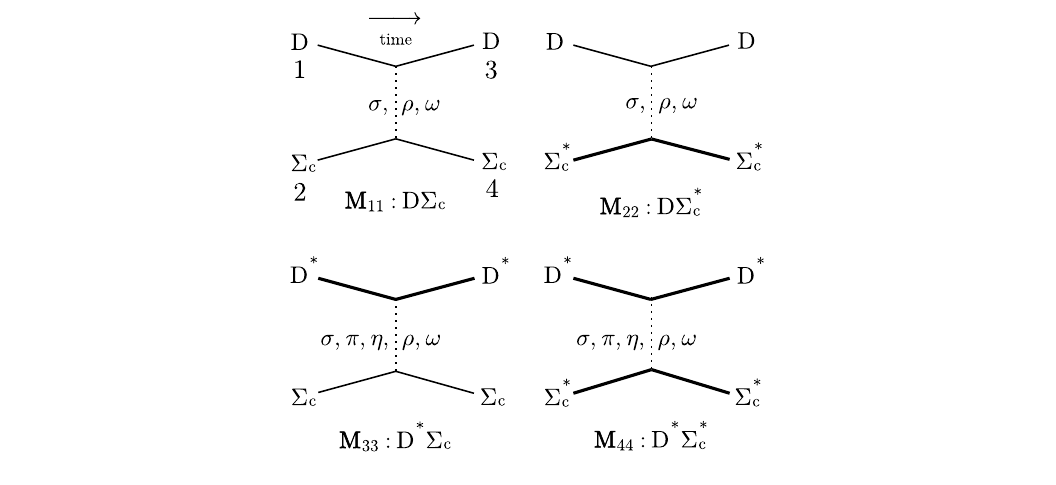}
 \caption{Tree-level Feynman diagrams for the four scattering processes $D^{(*)}\Sigma_c^{(*)} \to D^{(*)}\Sigma_c^{(*)}$.} \label{fig:FM_Graph}
\end{figure}

To give prominence to the isospin-breaking effects, we do not consider the
coupled-channel effects in the flavor space, but the mixing between different
partial waves. Thus, there are four systems for $D^{(*)}\Sigma_c^{(*)}$ in the
flavor space, $D \Sigma_{c}, D \Sigma_{c}^{*}, D^{*} \Sigma_{c}$, and $D^{*}
 \Sigma_{c}^{*}$, which are marked as $11, 22, 33$, and $44$ respectively, as
shown in Fig.~\ref{fig:FM_Graph} for the Feynman diagrams at tree level.

For each $D^{(*)}\Sigma_c^{(*)}$ system, we consider both the ground state
($S$-wave) configuration and its mixing with higher partial waves.
Specifically, we investigate the following spin-parity assignments: $J^P =
 1/2^-$ for $D\Sigma_c$; $J^P = 3/2^-$ for $D\Sigma_c^*$; $J^P = 1/2^-, 3/2^-$
for $D^*\Sigma_c$; and $J^P = 1/2^-, 3/2^-, 5/2^-$ for $D^*\Sigma_c^*$. The
partial-wave compositions considered for each spin-parity state are summarized
in Table~\ref{tab: SDstates}.

\begin{table}[h]
 \caption{
 The partial waves, denoted by the spectroscopic notation ${}^{2S+1}L_J$, for each spin-parity state are considered. We truncate the partial-wave expansion at the $D$-wave ($l = 2$), as the contribution from higher partial waves is expected to be negligible.}\label{tab: SDstates}
 \begin{ruledtabular}
 \begin{tabular}{c|cccc}
 $J^{P}$ & $\rm D \Sigma_{c}$ & $\rm D \Sigma_{c}^{*}$ & $\rm D^{*} \Sigma_{c}$ & $\rm D^{*} \Sigma_{c}^{*}$ \\
 \hline
 $\frac{1}{2}^{-}$ & $\ket{^{2} S_{1/2}}$ & $\times$ & $
\begin{pmatrix}
 \ket{^{2} S_{1/2}} \\
 \ket{^{4} D_{1/2}}
\end{pmatrix}$
 & $
 \begin{pmatrix}
 \ket{^{2} S_{1/2}} \\
 \ket{^{4} D_{1/2}} \\
 \ket{^{6} D_{1/2}}
\end{pmatrix}$ \\
 \hline
 $\frac{3}{2}^{-}$ & $\times$ & $
\begin{pmatrix}
 \ket{^{4} S_{3/2}} \\
 \ket{^{4} D_{3/2}}
\end{pmatrix}$
 & $
\begin{pmatrix}
 \ket{^{4} S_{3/2}} \\
 \ket{^{2} D_{3/2}} \\
 \ket{^{4} D_{3/2}}
\end{pmatrix}$
 & $\begin{pmatrix}
 \ket{^{4} S_{3/2}} \\
 \ket{^{2} D_{3/2}} \\
 \ket{^{4} D_{3/2}} \\
 \ket{^{6} D_{3/2}}
 \end{pmatrix}$ \\
 \hline
 $\frac{5}{2}^{-}$ & $\times$ & $\times$ & $\times$ & $
\begin{pmatrix}
 \ket{^{6} S_{5/2}} \\
 \ket{^{2} D_{5/2}} \\
 \ket{^{4} D_{5/2}} \\
 \ket{^{6} D_{5/2}}
\end{pmatrix}$

 \end{tabular}
 \end{ruledtabular}
\end{table}

Under isospin symmetry, the total isospin of $D^{(*)}\Sigma_c^{(*)}$ could be
$I = 1/2$ or $3/2$. In the isospin-$1/2$ case, there are two degenerate states,
$\ket{1/2, \pm1/2}$, while the isospin-$3/2$ case yields four degenerate
states, $\ket{3/2, \pm 3/2}$ and $\ket{3/2, \pm 1/2}$. However, once the
specific isospin breaking is considered, the degeneration is dissolved. In the
following, we present the explicit isospin wave functions for the $D\Sigma_c$
system. For isospin-$3/2$,
\begin{align}
 \ket{3/2,3/2} & =\ket{D^{+} \Sigma _{c}^{++}} , \nonumber \\
 \ket{3/2,1/2} & =\sqrt{\tfrac{2}{3}}\ket{D^{+} \Sigma _{c}^{+}} -\sqrt{\tfrac{1}{3}}\ket{D^0 \Sigma _{c}^{++}}, \nonumber \\
 \ket{3/2, -1/2} & =\sqrt{\tfrac{1}{3}}\ket{D^{+} \Sigma _{c}^0} -\sqrt{\tfrac{2}{3}}\ket{D^0 \Sigma _{c}^{+}}, \nonumber \\
 \ket{3/2, -3/2} & = - \ket{D^0 \Sigma _{c}^0},\label{isospin3/2}
\end{align}
while for isospin-$1/2$
\begin{align}
 \ket{1/2, 1/2} & =\sqrt{\tfrac{1}{3}}\ket{D^+ \Sigma _{c}^+} + \sqrt{\tfrac{2}{3}}\ket{D^0 \Sigma _{c}^{++}}, \nonumber \\
 \ket{1/2, -1/2} & =\sqrt{\tfrac{2}{3}}\ket{D^+ \Sigma _{c}^0} + \sqrt{\tfrac{1}{3}}\ket{D^0 \Sigma _{c}^+}.\label{isospin1/2}
\end{align}
The isospin wave functions for other $D^{(*)}\Sigma_c^{(*)}$ systems are similar.
From Eqs.~(\ref{isospin3/2}-\ref{isospin1/2}), it is evident that the Coulomb force contributes only to the charged isospin configurations, specifically the $\ket{3/2, 3/2}$, $\ket{3/2, 1/2}$, and $\ket{1/2, 1/2}$ states. The remaining states are neutral and thus receive no Coulomb contribution, as they involve either a $D^{0}$ or $\Sigma_c^{0}$.

\subsection{The Effective Strong Interaction}

We derive the strong interaction between a charmed $D^{(*)}$ and a charmed
$\Sigma_c^{(*)}$ within the OBE potential model. We start from the Lagrangians
satisfying the heavy quark symmetry and light SU(3)-flavor
symmetry ~\cite{Yan:1992gz, Wise:1992hn, Burdman:1992gh, Casalbuoni:1996pg,
 Falk:1992cx, Liu:2011xc},
\begin{eqnarray}
 \mathscr{L}_{\rm{eff}} = \mathscr{L}_{\mathbb{M}} + \mathscr{L}_{\mathbb{B}},
\end{eqnarray}
in which $\mathscr{L}_{\mathbb{M}}$ is for the mesonic part while $\mathscr{L}_{\mathbb{B}}$
is for the baryonic part. Specifically,
\begin{eqnarray}
 \mathscr{L}_{\mathbb{M}}
 &=& g_{\rm S} \langle H_{m} \sigma \bar{H}_{m} \rangle
 + {\rm i} \beta \langle H_{n} v_{\rho} (V^{\rho} - \mathbb{V}^{\prime \rho })_{nm}\bar{H}_{m} \rangle \nonumber \\
 &+&
 {\rm i} g \langle H_{n} X_{nm}^{\rho} \gamma_{\rho} \gamma_{5}\bar{H}_{m} \rangle
 + {\rm i} \lambda \langle H_{n} F_{nm}^{\alpha \beta} \sigma_{\alpha \beta} \bar{H}_{m} \rangle, \label{Lagrangianmeson}
\end{eqnarray}
and
\begin{eqnarray}
 \mathscr{L}_{\mathbb{B}}
 &=&
 l_{\rm S} g_{\mu \nu } {\rm Tr} [\bar{S}^{\mu} \sigma S^{\nu}]
 + {\rm i} \beta_{\rm S} g_{\mu \nu} {\rm Tr} [\bar{S}^{\mu} (V^{\rho} - \mathbb{V}^{\prime \rho}) v_{\rho} S^{\nu} ] \nonumber \\
 &+& \tfrac{3}{2} g_{1} \varepsilon_{\mu \nu \alpha \beta} v^{\beta} {\rm Tr} [\bar{S}^{\mu} X^{\alpha} S^{\nu} ]
 +\lambda_{\rm S} {\rm Tr} [\bar{S}^{\mu } F_{\mu \nu } S^{\nu } ], \label{Lagrangianbaryon}
\end{eqnarray}
where $H$ is a superfield formed by the heavy pseudoscalar and vector mesons under the heavy quark spin symmetry (HQSS).
The specific expression of $H$ is
\begin{eqnarray}
 H_a &\equiv& \frac{1+\not{v}}{2} (\gamma_{\mu } P_{a}^{*\mu } - \gamma_{5} P_{a}),
\end{eqnarray}
while the conjugate field $\bar{H}$ is defined as
\begin{eqnarray}
 \bar{H}_a \equiv \gamma ^{0} H_{a}^{\dagger} \gamma ^{0} =(\gamma _{\mu } P_{a}^{*\mu \dagger } +\gamma _{5} P_{a}^{\dagger } )\frac{1+\not{v}}{2}.
\end{eqnarray}
Here,
\begin{equation}
 P^{(*)} = \left(D^{(*)0} , D^{(*)+} , D^{(*)}_{s}\right)^{\rm T}
\end{equation}
is the vector formed by the heavy pseudoscalar (vector) mesons under
the light SU(3)-flavor and heavy-flavor symmetries.
Similarly, the spin-$1/2$ and spin-$3/2$ charmed baryons can also form a superfield $S^{\mu}$ within HQSS as
\begin{eqnarray}
 S^{\mu } &\equiv& -\frac{1}{\sqrt{3}} (\gamma ^{\mu } +v^{\mu } )\gamma ^{5}B_{6} + B_{6}^{*\mu },
\end{eqnarray}
while the conjugate field $\bar{S}^\mu$ is defined as
\begin{eqnarray}
 \bar{S}^{\mu } &\equiv& S^{\mu \dagger } \gamma ^{0} = \frac{1}{\sqrt{3}}\bar{B}_{6} \gamma ^{5} (\gamma ^{\mu } +v^{\mu } )+\bar{B}_{6}^{*\mu }.
\end{eqnarray}
The $B_6 (B^{*\mu}_6)$ represents the matrix formed by the spin-$1/2$ ($3/2$) baryons which have the configuration of one heavy and two
light quarks,
\begin{eqnarray}
 B_{6} \equiv
 \begin{bmatrix}
 \rm{\Sigma}^{ ++} & \tfrac{1}{\sqrt{2}}\rm{\Sigma }^{ +} & \tfrac{1}{\sqrt{2}}\rm{\Xi }^{\prime +} \\
 \tfrac{1}{\sqrt{2}}\rm{\Sigma }^{ +} & \rm{\Sigma }^{0} & \tfrac{1}{\sqrt{2}}\rm{\Xi }^{\prime 0} \\
 \tfrac{1}{\sqrt{2}}\rm{\Xi }^{\prime +} & \tfrac{1}{\sqrt{2}}\rm{\Xi }^{\prime 0} & \Omega^{ 0}
 \end{bmatrix}.
\end{eqnarray}

In the above expressions, $v^\mu$ denotes the 4-velocity of the heavy meson or
baryon, satisfying $v_\mu v^\mu = 1$. In the heavy quark limit ($m_Q \to
 \infty$), we adopt the static limit $v^\mu = (1, \vec{0})$. The symbol $\langle
 \ldots \rangle$ represents the trace over Dirac $\gamma$ matrices, while
$\mathrm{Tr}[\ldots]$ denotes the trace over flavor indices. The quantities
$V^\mu$ and $X^\mu$ correspond to the vector and axial-vector currents
respectively, constructed from the light pseudoscalar mesons,
\begin{eqnarray}
 V^{\mu} &\equiv& \frac{1}{2} (\xi^{\dagger } \partial ^{\mu} \xi + \xi \partial ^{\mu} \xi^{\dagger } ), \nonumber \\
 X^{\mu} &\equiv& \frac{1}{2} (\xi^{\dagger } \partial ^{\mu} \xi - \xi\partial ^{\mu} \xi^{\dagger } ),
\end{eqnarray}
where $\xi$ is defined as
\begin{eqnarray}
 \xi \equiv \exp ({\rm i} \mathbb{P} / f_{\pi} )
\end{eqnarray}
with $\mathbb{P}$ the matrix of the pseudoscalar mesons,
\begin{eqnarray}
 \mathbb{P} &\equiv&
 \begin{bmatrix}
 \tfrac{\pi^{0}}{\sqrt{2}} +\tfrac{\eta^{0}}{\sqrt{6}} & \pi^{+} & \rm{K^{+}} \\
 \pi^{-} & -\tfrac{\pi^{0}}{\sqrt{2}} +\tfrac{\eta ^{0}}{\sqrt{6}} & \rm{K}^{0} \\
 \rm{K}^{-} & \rm{\bar{K}}^{0} & - \tfrac{2}{\sqrt{6}} \eta^{0}
 \end{bmatrix},
\end{eqnarray}
and $f_{\pi} = 0.132$~GeV the pion decay constant. In Eqs.~\eqref{Lagrangianmeson}-\eqref{Lagrangianbaryon},
$F^{\mu \nu}$ is the tensor current constructed by the light vector mesons,
\begin{eqnarray}
 F^{\mu \nu} \equiv \partial^{\mu } \mathbb{V}^{\prime \nu} - \partial ^{\nu} \mathbb{V}^{\prime \mu } + [\mathbb{V} ^{\prime \mu } , \mathbb{V}^{\prime \nu } ],
\end{eqnarray}
where
\begin{eqnarray}
 \mathbb{V}^{\prime \mu} \equiv {\rm i} g_{\rm V} \mathbb{V}^{\mu} / \sqrt{2},
\end{eqnarray}
with $g_{V} = 5.83$ and the matrix for the light vector mesons,
\begin{eqnarray}
 \mathbb{V}^{\mu} &\equiv&
 \begin{bmatrix}
 \tfrac{\rho^{0}}{\sqrt{2}} +\tfrac{\omega^{0}}{\sqrt{2}} & \rho^{+}& \rm{K^{*+}} \\
 \rho^{-} & -\tfrac{\rho^{0}}{\sqrt{2}} +\tfrac{\omega^{0}}{\sqrt{2}} & \rm{K}^{*0} \\
 \rm{K}^{*-} & \rm{\bar{K}}^{*0} & \phi^{0}
 \end{bmatrix}^{\mu}.
\end{eqnarray}
In Table~\ref{tab:constants}, we provide the values for other coupling constants appearing in the Lagrangians ~\cite{ChenRui_2021}.
\begin{table}[h]
 \caption{\label{tab:constants}
 The values of coupling constants used in the calculation.}
 \begin{ruledtabular}
 \begin{tabular}{cccc}
 $g_{\rm S}$ & $\beta$ & $g$ & $\lambda$ \\
 0.76& $-0.90$ & 0.59 & $-0.56$ ${\rm GeV}^{-1}$ \\
 \hline
 \addlinespace
 $l_{\rm S}$ & $\beta_{\rm S}$ & $g_1$ & $\lambda_{\rm S}$ \\
 6.20& 1.74 & 0.94 & 3.31 ${\rm GeV}^{-1}$
 \end{tabular}
 \end{ruledtabular}
\end{table}

To obtain the effective potentials in a non-relativistic limit, we apply the
Breit approximation
\begin{eqnarray}
 V =-\frac{\mathcal{M}} {\sqrt{\prod (2m_{i} )}\sqrt{\prod (2m_{f} )}},
\end{eqnarray}
in which $m_{i}(m_{f})$ is the mass of the incoming (outgoing) particle. The scattering amplitude $\mathcal{M}$ can be obtained by calculating the
Feynman diagrams in Fig.~\ref{fig:FM_Graph}. Through straightforward calculation, we obtain the effective potential $V(q)$ in momentum space. The corresponding coordinate-space potential is then derived via the Fourier transformation
\begin{eqnarray}
 V(r) = \hat{\mathcal{F}}\left[V({\bf q})\right] = \int \frac{ d^3{\bf q}}{(2\pi)^3} e^{{\rm i}{\bf q} \cdot {\bf r}} V({\bf q}).
\end{eqnarray}

It is found that the effective potentials are classified into three types,
i.e., the center force, spin-orbital force and tensor force. The latter two are
well known to play crucial roles in nuclear physics. The spin-orbit force is
essential for explaining the magic numbers in atomic nuclei, while the tensor
force mixing the partial waves with $l$ and $l\pm2$ plays a key role in
describing the weakly bound deuteron. Overall, the effective potential can be
written in the following general form,
\begin{eqnarray}
 V(r) = f_{\rm c}(r) \mathcal{O}_{\rm cen} + f_{\rm SL}(r)\mathcal{O}_{\rm SL} + f_{\rm ten}(r)\mathcal{O}_{\rm ten},
\end{eqnarray}
where $f_i(r)$ is a scalar function while $\mathcal{O}_i$ is the operator for each type of interactions.

However, the effective potential for exchanging a light pseudoscalar/vector
meson is singular due to the following results,
\begin{eqnarray}
 \hat{\mathcal{F}} \left[\frac{{\bf q}^2}{{\bf q}^{2} +u^{2}}\right] = - 4\pi \delta ({\bf r}) + f(r, u).
\end{eqnarray}
From a physical perspective, hadrons are not elementary particles but possess nontrivial internal structure. To account for this in theoretical calculations, a form factor is typically introduced as
\begin{eqnarray}
 F_{n}(\Lambda, m, q) &=& \left(\frac{\Lambda^{2} - m^{2}}{\Lambda^{2} - q^{2}} \right)^{n}, \label{formfactor}
\end{eqnarray}
which is called $n$-point form factor. In Eq.~\eqref{formfactor}, $q$ and $m$ are the transferred 4-momentum and mass of the exchanged meson respectively, and $\Lambda$ is the momentum cutoff introduced to suppress the high-momentum contribution. 
In such a way, the singular behavior of the interaction potential can be regularized when two hadrons approach to each other. In our calculation, we apply the
monopole ($n = 1$) form factor,
\begin{eqnarray}
 V(r) = \int \frac{{\rm d}^3 {\bf q}}{(2\pi)^3} e^{{\rm i} {\bf q} \cdot {\bf r}} V({\bf q})F_1^{2} (\Lambda, m, q).
\end{eqnarray}
Note that $F_1(\Lambda, m, q) \to 1$ as $\Lambda \to \infty$, in which limit the form factor becomes trivial and the hadron is effectively treated as a point-like particle. However, the appropriate value of $\Lambda$ for a given calculation is not a priori known. In practice, one typically adopts values in the range $1.0 < \Lambda < 1.5$~GeV, which have been shown to accurately describe both the properties of the weakly bound deuteron and neutron-proton scattering within the framework of OBE model. In the present work, we therefore adopt this range as the physically motivated choice for $\Lambda$.

\subsection{Isospin Breaking}
Essentially, isospin breaking originates from two fundamental sources: the
strong interaction, arising from the mass difference between up and down
quarks, and the electromagnetic interaction, arising from their charge
difference. At the hadronic level, these effects manifest in several ways in
the present study. First, the strong interaction contribution to isospin
breaking is reflected in the mass differences among hadrons belonging to the
same isospin multiplet--specifically, between $D^{(*)0}$ and $D^{(*)+}$,
between $\pi^0$ and $\pi^\pm$, and among the $\Sigma_c^{(*)}$ states. Second,
electromagnetic isospin breaking enters through the Coulomb interaction between
charged $D^{(*)}$ and $\Sigma_c^{(*)}$ states. In principle, isospin breaking
also affects the coupling constants for mesons with different $I_3$ projections
within the same isospin multiplet. However, such effects are expected to be
minor and are neglected in this work. Accordingly, we retain the
SU(3)-flavor-symmetric Lagrangians introduced earlier. It is straightforward to
incorporate the isospin breaking from the strong interaction. In the following,
we will focus on the electromagnetic interaction, i.e. the coulomb force
between the charged $D^{(*)}$ and $\Sigma_c^{(*)}$.

Recall that the Coulomb interaction between two charged point-like particles is
\begin{eqnarray}
 V_{\rm C}(r) = \frac{1}{\alpha}\frac{Z_1 Z_2}{r} \equiv N k \frac{1}{r} ,\label{coulomb}
\end{eqnarray}
where $k = 1/\alpha$ is the Coulomb constant and $N = Z_1 Z_2$, 
which can not be applied directly to the hadrons with charge distributions.
Due to the lack of experimental data, the charge distributions for most hadrons are not well determined. Nevertheless, the Coulomb interaction between charged hadrons can be modeled approximately using an exponential charge distribution. Alternatively, one may derive the Coulomb potential by employing an $n$-point form factor. The relationship between these two approaches is subtle. To clarify this, we derive the Coulomb force using both methods and compare the resulting potentials.

To compute the Coulomb interaction using the $n$-point form factor, we first rewrite the Coulomb force between two point-like charge particles as
\begin{align}
 {V}_{\rm C}(r) = N k \frac{e^{-\varepsilon r}}{r} , \quad \varepsilon \to 0, \label{coulomb_point}
\end{align}
then perform an inverse Fourier transform of Eq.~(\ref{coulomb_point}) to obtain the
Coulomb potential in momentum space ~\cite{Goitein:1967kaq, AUERBACH:1972yql}
\begin{align}
 {\tilde V}_{\rm C}({\bf q}) =\hat{\mathcal{F}}^{-1}[V_{\rm C}(r)] = N k \frac{4\pi }{{\bf q}^{2} + \varepsilon ^{2}} \xrightarrow[]{\varepsilon \to 0} Nk\frac{4\pi }{{\bf q}^{2}}.
\end{align}

Incorporating the $n$-point form factor $F_{n}(\eta, {\bf q})$ and performing the Fourier transform yields the modified Coulomb interaction:
\begin{align}
 V^{2n}_{\rm C}(r) & = \hat{\mathcal{F}} \left[ F_n(\eta, {\bf q}) {\tilde V}_{\rm C}({\bf q}) \right]\nonumber \\
& = N k \frac{1}{r} \left[ 1-G_{2n}( \eta r) e^{-\eta r} \right], \label{Formed Coulomb}
\end{align}
where ~\cite{Goitein:1967kaq}
\begin{eqnarray}
 F_{n}(\eta, {\bf q}) = \left(\frac{\eta^{2}}{\eta^{2} + {\bf q}^{2}} \right)^{n} \label{CL Form factor}
\end{eqnarray}
is the $n$-point form factor, and 
$G_{n}(x)$ is the $n$-th form polynomial
\begin{align}
 \begin{aligned}
 G_{0}( x) & =0, \\
 G_{1}( x) & =1, \\
 G_{2}( x) & =1+\frac{1}{2} x,\\
 G_{3}( x) & =1+\frac{5}{8} x+\frac{1}{8} x^{2}, \\
 G_{4}( x) & =1+\frac{11}{16} x+\frac{3}{16} x^{2} +\frac{1}{48} x^{3}, \\
 \cdots . \label{Form polynomial}
 \end{aligned}
\end{align}

On the other hand, given that a hadron state has sizable structure, an exponential charge distribution can be introduced for a charged hadron,
\begin{eqnarray}
 \rho_{i}(r) = Q_{i} \frac{a^3}{8 \pi} e^{-a r}, \quad (i=1,2), \label{Exp distribution}
\end{eqnarray}
in which $Q_{i}$ is the charge of the $i$-th hadron. The parameter is typically taken as $a = 2\sqrt{6} / r_{\rm{Q}}$~\cite{Lyu:2021qsh}, where $ r_{\rm{Q}} $ denotes the charge radius of the hadron.

Then, one can calculate the classical electromagnetic potential as
\begin{eqnarray}
 V_{\rm C}( r) &=& k\int \frac{\rho _{1} (|\vec{r}_{1} |)\rho _{2} (|\vec{r} -\vec{r}_{2} |)}{|\vec{r}_{1} -\vec{r}_{2} |} {\rm d} \vec{r}_{1} {\rm d} \vec{r}_{2} \nonumber \\
 &=& Nk \frac{1}{r} \left[ 1 - G_{4}(ar) e^{-ar} \right], \label{CL4}
\end{eqnarray}
where this $G_{4}$ function is defined the same as in Eq.~(\ref{Form polynomial}).

It's clear that the integral result of
Eq.~(\ref{CL4}) is exactly $V_{\rm C}^{4}(r)$ in Eq.~(\ref{Formed Coulomb}), and the parameter $a$ in Eq.~(\ref{Exp distribution}) is just the cutoff $\eta$ in Eq.~(\ref{CL Form factor}).
This result denotes that the two methods - either by applying an $n$-point form factor, or by applying an exponential charge distribution on the original Coulomb interanction, is equivalent. Besides, $V_{\rm C}^2(r)$ is the potential calculated by applying an exponential charge distribution for one hadron and a point-like charged hadron,
while $V_{\rm C}^0(r)$ represents just the original Coulomb interaction between two point-like charged particles as Eq.~(\ref{coulomb}).

The charge radius $r_{\rm Q}$ is not generally known for most hadrons. In practice, one often adopts the approximate relation $r_{\rm{Q}} \approx \sqrt{6} / \Lambda$, where $\Lambda$ is the momentum cutoff for the strong interaction in OBE model. This leads to the simple estimate $a \approx 2\Lambda$.

\begin{figure}[htp]
 \includegraphics[width=0.5\textwidth]{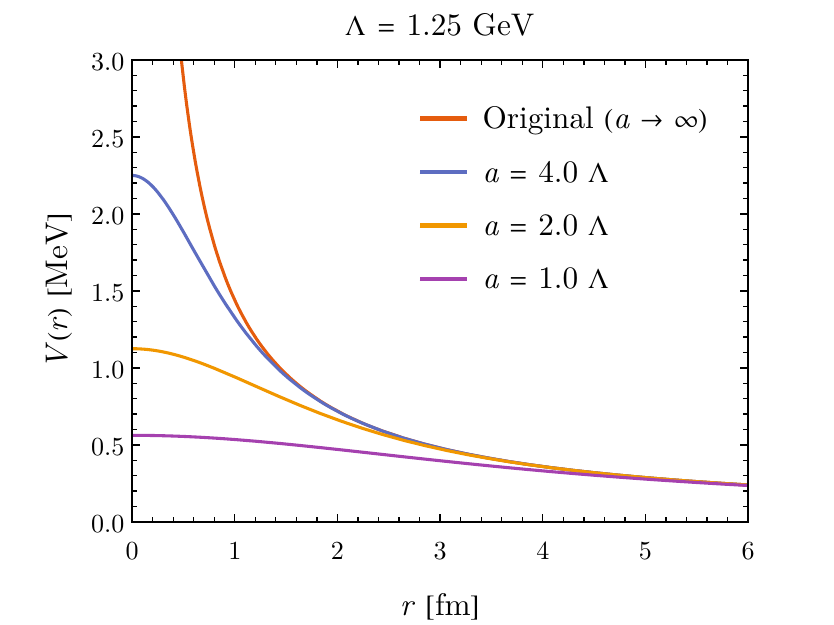}
 \caption{(Color online) Coulomb interaction between two charged point-like particles and
 the ones between charged hadrons calculated using Eq.~(\ref{CL4}). For the latter, three cutoff values are adopted as $a = 1.0\Lambda,~ 2.0\Lambda, ~4.0\Lambda$, with $\Lambda=1.25$ GeV. }\label{fig:VCL_Plot}
\end{figure}

In Fig.~\ref{fig:VCL_Plot}, we present the Coulomb potential calculated from
Eq.~(\ref{CL4}), together with the point-like Coulomb potential for comparison.
To examine the dependence on the cutoff parameter $a$, we consider three
values, $a =1.0 \Lambda$, $2.0\Lambda$, and $4.0\Lambda$, with a typical $\Lambda = 1.25$~GeV. The results show that the Coulomb potential exhibits only a weak
dependence on $a$ for $r > 1.0$~fm. However, the cutoff plays an essential role
in regularizing the $r \to 0$ divergence inherent to the point-like
interaction.

\subsection{Schr\"odinger Equation with Isospin Breaking}

Since the interaction between two particles depends solely on their relative
distance, a two-body system has spherical symmetry, i.e. one can separate the
radial part of the wave function from its angular part. After integrating out
the angular degrees of freedom, the radial Schr\"odinger equation takes the
form
\begin{eqnarray}
 \left[ -\frac{1}{2\mu } \partial {_{r}}^{2} +\frac{l( l+1)}{2\mu r^{2}} +V(r) \right] u(r) = Eu(r),
\end{eqnarray}
with $\mu = m_1 m_2 / (m_1 + m_2)$ the reduced mass and $u(r) = rR(r)$ the modified radial wave function.
$l$ and $V(r)$ represent the quantum numbers of orbital angular momentum and the interaction potential, respectively.
For the case we include the isospin breaking, i.e., the Coulomb interaction between charged hadrons,
the threshold difference of $D^{(*)}\Sigma_c^{(*)}$ belonging the same isospin multiplet,
the matrix element of Hamiltonian is expressed as
\begin{eqnarray}
 H_{pq} = \bra{\mathcal{P}}\hat{H}\ket{\mathcal{Q}} = T_{pq} +V_{pq} +\Delta M_{pq},
\end{eqnarray}
where $\mathcal{P}$ and $\mathcal{Q}$ are two partial wave channels. The kinetic energy is always diagonal and
can be written as
\begin{eqnarray}
 T_{pq} = -\frac{1}{2\mu _{p}} \partial_{r}^{2} \delta _{pq} +\frac{l_{p}( l_{p} +1)}{2\mu _{p} r^{2}} \delta _{pq},
\end{eqnarray}
where $\mu_{p}$ and $l_{p}$ are the reduced mass and quantum number of orbital angular momentum in the
$\mathcal{P}$ channel, respectively. The second term is the centrifugal potential, which is positive and grows with increasing $l_p$.

The effective potential includes the strong interaction from exchanging a light
meson and the Coulomb interaction,
\begin{align}
 V_{pq}(r) & = V^{\rm OBE}_{pq}(r) + V^{\rm C}_{pq}(r).
\end{align}
To illustrate the construction of the strong interaction potential matrix, we consider the $D^*\Sigma_c$ system in the isospin state $\ket{I,I_3} = \ket{1/2, 1/2}$. In the basis ${D^{*+}\Sigma_c^+, D^{*0}\Sigma_c^{++}}$, i.e., the strong interaction potential matrix is
expressed as
\begin{align}
 V^{\rm OBE}(r) & =
 \begin{bmatrix}
 V^{\rm OBE}_{11} (r)
 & V^{\rm OBE}_{12}(r) \\
 V^{\rm OBE}_{21}(r)
 & V^{\rm OBE}_{22}(r)
 \end{bmatrix}, \label{potential_matrix}
\end{align}
where
\begin{align}
 V^{\rm OBE}_{11}(r) & = \bra{D^{*+} \Sigma _{c}^{+}} {\hat V}^{\rm OBE}(r) \ket{D^{*+} \Sigma _{c}^{+}}, \nonumber \\
 V^{\rm OBE}_{12}(r) & = \bra{D^{*+} \Sigma _{c}^{+}} {\hat V}^{\rm OBE}(r) \ket{D^{*0} \Sigma _{c}^{++}}, \nonumber \\
 V^{\rm OBE}_{22}(r) & = \bra{D^{*0} \Sigma _{c}^{++}} {\hat V}^{\rm OBE}(r) \ket{D^{*0} \Sigma _{c}^{++}}. \nonumber
\end{align}
The matrix of Coulomb interaction is expressed as
\begin{align}
 V^{\rm C}(r) & =
 \begin{bmatrix}
 V^{\rm C}_{11}(r) & 0 \\
 0 & 0
 \end{bmatrix},
\end{align}
with $V^{\rm C}_{11}(r) = \bra{D^{*+}\Sigma_c^+} {\hat V}^{\rm C}(r) \ket{D^{*+} \Sigma _{c}^{+}}$. Since $D^{*0}$ is neutral, all remaining matrix elements of $V^{\rm C}(r)$ vanish.

The threshold difference is given by
\begin{align}
 \Delta M_{pq} = \delta M_{p} \delta _{pq}, \quad \delta M_{p} = M_{p} - M_{\rm min},
\end{align}
where $M_{p}$ is the mass threshold of the $p$-th channel, and $M_{\rm min}$ is the lowest threshold among all considered flavor channels.

When mixing between different partial waves is taken into account, each isospin
state generally encompasses multiple partial-wave components. For instance, the
$D^{*}\Sigma_{c}$ system with $\ket{I, I_3} = \ket{1/2, 1/2}$ and $J^P = 1/2^-$
involves two partial waves: ${}^{2}S_{1/2}$ and ${}^{4}D_{1/2}$. In such cases,
each element of the potential matrix becomes a $2 \times 2$ submatrix in the
partial-wave bases.

Alternatively, one can also write the potential matrix in the isospin bases,
which is easily obtained from~Eq.~(\ref{potential_matrix}) via the following
orthogonal transformation,

\begin{align}
 V^{\rm iso}(r) = U V(r) U^{-1},\label{transform}
\end{align}
where $U$ is the transform matrix. For the state with $I_3 = 1/2$,
\begin{align}
 U = \left(
 \begin{array}{ c c }
 \sqrt{\tfrac{1}{3}} & -\sqrt{\tfrac{2}{3}} \\
 \sqrt{\tfrac{2}{3}} & \sqrt{\tfrac{1}{3}}
 \end{array}
 \right), \quad
 U^{-1} = U^{\rm T}.
\end{align}

Firstly, we write the specific expression of potential matrix
Eq.~(\ref{potential_matrix}) as
\begin{align}
 V(r) & = V_{\sigma }(r) \hat{\bf I} + V_{8}(r), \\
 V_{8}(r) & =
 \begin{bmatrix}
 \tfrac{V_\eta(r) +3V_{\omega }(r)}{6} &
 -\frac{V_{\pi }(r) +V_{\rho }(r)}{\sqrt{2}} \\
 -\frac{V_{\pi }(r) +V_{\rho }(r)}{\sqrt{2}} &
 \frac{V_{\eta }(r) + 3V_{\omega }(r) + 3V_{\pi }(r) +V_{\rho }(r)}{6}
 \end{bmatrix},
\end{align}
where $V_{\sigma, \pi, \rho, \eta, \omega}$ is the potential for
exchanging a $\sigma, \pi, \rho, \eta$, or $\omega$ meson. Performing the transformation as in Eq.~(\ref{transform}), the
potential matrix converts into
\begin{align}
 V^{\rm iso}(r) = & V_{\sigma }(r) \hat{\bf I} + V_{8}^{\rm iso}(r), \\
 V_{8}^{\rm iso}(r) = & {\bf Diag} \big[\left(\tfrac{1}{6} V_{\eta }(r) +\tfrac{1}{2} V_{\omega }(r)\right) +( V_{\pi }(r) +V_{\rho }(r)), \nonumber \\
 & \left(\tfrac{1}{6} V_{\eta }(r) +\tfrac{1}{2} V_{\omega }(r)\right) -\tfrac{1}{2}( V_{\pi }(r) +V_{\rho }(r)) \big],
\end{align}
or in a more compact form
\begin{align}
 V^{\rm iso}(r) =
 \begin{bmatrix}
 V^{D^{*} \Sigma _{c}}_{11}(r) & 0 \\
 0 & V^{ D^{*} \Sigma _{c}}_{22}(r)
 \end{bmatrix},
\end{align}
in which
\begin{align}
 V^{D^*\Sigma_c}_{11}(r) & = \bra{1/2, 1/2} {\hat V}^{D^*\Sigma_c}(r) \ket{1/2, 1/2}, \nonumber \\
 V^{D^*\Sigma_c}_{22}(r) & = \bra{3/2, 1/2} {\hat V}^{D^*\Sigma_c}(r) \ket{3/2, 1/2}.
\end{align}
It is clear that the isospin-$1/2$ and -$3/2$ states will decouple in the isospin symmetry limit.
Applying the same transformation to the two isospin-breaking terms yields
\begin{align}
 V_{\rm C}^{\rm iso}(r) = U V_{\rm C}(r) U^{-1} =
 \begin{pmatrix}
 \frac{1}{3}& \frac{\sqrt{2}}{3} \\
 \frac{\sqrt{2}}{3} & \frac{2}{3}
 \end{pmatrix} V_{\rm C}(r), \label{coulomb_couple}
\end{align}
and
\begin{align}
 \Delta M^{\rm iso} = U \Delta M U^{-1} =
 \begin{pmatrix}
 \frac{1}{3}& \frac{\sqrt{2}}{3} \\
 \frac{\sqrt{2}}{3} & \frac{2}{3}
 \end{pmatrix} \Delta M. \label{massdifference_couple}
\end{align}
It is now evident from Eqs.~\eqref{coulomb_couple}-\eqref{massdifference_couple} that the presence of isospin-breaking terms induces mixing between the $I = 1/2$ and $I = 3/2$ states. Consequently, total isospin $I$ is no longer a good quantum number, although its third component $I_3$ remains conserved.

\section{Numerical Results}\label{numerical_results}

In this section, we present our numerical results and quantitative analyses. We
first show the results obtained under isospin symmetry, followed by those with
isospin breaking explicitly included. The effects of isospin breaking can then
be assessed by comparing the two sets of results.

\subsection{The $D \Sigma_{c}$ systems}\label{sec:A}
The $S$-wave ground state of the $D\Sigma_c$ system carries spin-parity $J^P = 1/2^-$, involving only the partial wave ${}^{2}S_{1/2}$. Its total isospin can be either $I = 1/2$ or $3/2$. At tree level, the absence of $D\mathcal{P}D$ interaction vertex forbids coupling to diagrams involving a light pseudoscalar meson $\mathcal{P}$ (e.g., a pion) exchanged between the $D$ and $\Sigma_c$. Consequently, within the OBE potential model, the interaction proceeds solely via the exchange of heavier scalar and vector mesons, such as $\sigma$, $\rho$, and $\omega$, which generate the medium- and short-range forces.

As shown in Table~\ref{DSigma}, our numerical results indicate that the
isospin-$1/2$ $D\Sigma_c$ system can form a loosely bound state. Under the
isospin symmetry, the binding energy ranges from $0.36$ to $13.69$~MeV, and the
corresponding root-mean-square ({\it rms}) radius ranges from $4.75$ to $1.12$~fm,
for the momentum cutoff varied in the range $\Lambda = 1.14$ - $1.30$~GeV.
These binding properties suggest that this state can be regarded as a typical
hadronic molecule. In contrast, no bound state is found for the isospin-$3/2$
$D\Sigma_c$ system for cutoff values up to $5.0$~GeV.

When isospin breaking is included, both the binding energy and the {\it rms}
radius of the isospin-$1/2$ $D\Sigma_c$ state shift appreciably, indicating
sizable isospin-breaking effects in this hadronic molecular candidate. For
instance, at a cutoff value of $\Lambda = 1.25$~GeV, the {\it rms} radius
increases from $1.41$~fm to $1.44$~fm, while the binding energy decreases from
$7.67$~MeV to $5.90$~MeV upon the inclusion of isospin breaking. This
corresponds to a roughly $23\%$ correction to the binding energy.

\begin{table}
 \caption{Binding solutions of the $D \Sigma_{c}$ system. The two cases I and II mean without (w/o) and
 with (w/) isospin breaking. $\Lambda$, $E_b$, and $r_{rms}$ are the momentum cutoff
 (in units of GeV), binding energy (in units of MeV) and root-mean-square ({\it rms}) radius
 (in units of fm), respectively. $\times$ means no bound state when $\Lambda < 5.0$~GeV.
 The obtained bound states are in pure S-wave. } \label{DSigma}
 \begin{ruledtabular}
 \begin{tabular}{cc ccc}
 $\ket{I, I_3}$& Cases & $\Lambda$ (GeV) & $E_b$ (MeV) & $r_{rms}$ (fm) \\
 \midrule
 \multirow{6}{*}{$\ket{1/2, 1/2}$} & & $1.14$ & $0.36$ & $4.75$ \\
 & I (w/o ) & $1.25$ & $7.67$ & $1.41$ \\
 & & $1.30$ & $13.69$ & $1.12$ \\
 \cmidrule{2-5}
 & & $1.18$ & $0.41$ & $2.67$ \\
 & II (w/) & $1.25$ & $5.90$ & $1.44$ \\
 & & $1.30$ & $11.83$ & $1.14$ \\
 \midrule
 $\ket{3/2, 1/2}$ & & \multicolumn{3}{c}{$\times$} \\
 \midrule
 $\ket{3/2, 3/2}$ & & \multicolumn{3}{c}{$\times$} \\
 \midrule
 \end{tabular}
 \end{ruledtabular}
\end{table}

\subsection{The $D \Sigma_{c}^{*}$ systems}\label{DSigmastar}
The ground state of the $D\Sigma_c^*$ system with $J^P = 3/2^-$ involves two partial waves: ${}^4S_{3/2}$ and ${}^4D_{3/2}$. However, we find that the $D$-wave contribution is negligible in our calculation.

The numerical results of $D\Sigma_c^*$ are shown in Table~\ref{3/2DSigmastar}.
Similarly to the results of the $D\Sigma_c$ system, the isospin-$1/2$
$D\Sigma_c^*$ can form a loosely bound state. Under the isospin symmetry, the
binding energy and {\it rms} radius of this state are 0.45 to 14.27~MeV and
4.44 to 1.10~fm respectively, when the momentum cutoff varied in the range
$\Lambda = 1.14$ - $1.30$~GeV. However, we could not obtain any binding
solutions for the isospin-$3/2$ $D\Sigma_c^*$ system when the momentum cutoff
is smaller than $5.0$~GeV.

Taking into account the isospin breaking, the binding energy and {\it rms}
radius of the isospin-$1/2$ $D\Sigma_c^*$ state also exhibit noticeable shifts.
Specifically, at $\Lambda = 1.25$~GeV, the {\it rms} radius increases from
$1.38$~fm to $1.41$~fm, while the binding energy decreases from $8.08$~MeV to
$6.20$~MeV, indicating about $23\%$ isospin-breaking correction to the binding
energy.
\begin{table}
 \caption{Binding solutions of the $D \Sigma_{c}^{*}$ system. $\Lambda$, $E_b$ and $r_{rms}$ are the momentum cutoff
 (in units of GeV), binding energy (in units of MeV) and {\it rms} radius (in units of fm), respectively. It
 should be mentioned that the obtained bound state is pure S-wave. $\times$ mean no bound states for $\Lambda < 5.0$~GeV. }\label{3/2DSigmastar}
 \begin{ruledtabular}
 \begin{tabular}{ccccc}
 $\ket{I, I_3}$& Cases & $\Lambda$ (GeV) & $E_b$ (MeV) & $r_{rms}$ (fm) \\
 \midrule
 \multirow{6}{*}{$\ket{1/2, 1/2}$} & & $1.14$ & $0.45$ & $4.44$ \\
 & I (w/o) & $1.25$ & $8.08$ & $1.38$ \\
 & & $1.30$ & $14.27$ & $1.10$ \\
 \cmidrule{2-5}
 & & $1.18$ & $0.50$ & $2.53$ \\
 & II (w/) & $1.25$ & $6.20$ & $1.41$ \\
 & & $1.30$ & $12.29$ & $1.12$ \\
 \midrule
 $\ket{3/2, 1/2}$ & & \multicolumn{3}{c}{$\times$} \\
 \midrule
 $\ket{3/2, 3/2}$ & & \multicolumn{3}{c}{$\times$} \\
 \midrule
 \end{tabular}
 \end{ruledtabular}
\end{table}

\subsection{The $D^{*} \Sigma_{c}$ systems}\label{DstarSigma}

The spin-parity of the $D^*\Sigma_c$ system could be $J^P=1/2^-$ or $3/2^-$ in
the ground states. As illustrated in Table~\ref{tab: SDstates}, we shall
consider the ${}^2S_{1/2}$ and ${}^4D_{1/2}$ partial waves for the $J^P =
 1/2^-$ channel, while the ${}^4S_{3/2}$, ${}^2D_{3/2}$ and ${}^4D_{3/2}$
partial waves for the $J^P = 3/2^-$ channel. We present the numerical results
for the $D^*\Sigma_c$ systems with $J^P = 1/2^-$ and $3/2^-$ in
Table~\ref{1/2DstarSigma} and Table \ref{3/2DstarSigma}, respectively.

The isospin-$3/2$ $D^*\Sigma_c$ with $J^P = 1/2^-$ can also form a loosely
bound state with binding energy $0.22 - 18.77$~MeV and {\it rms} radius $5.11 -
 0.87$~fm for the momentum cutoff tuned to $1.40 - 1.90$~GeV, in the isospin
symmetry limit. It is predominantly $^2S_{1/2}$, with a probability larger than
$99\%$, and minor contribution from $^4D_{1/2}$, less than $1\%$. The state
$D^{*+}\Sigma_c^{++}$ ($\ket{I,I_3} = \ket{3/2, 3/2}$) is particularly
interesting since it has three positive unit charges and cannot be formed by
three quarks in the traditional way of the quark model. From this perspective,
it can be taken as an `absolute' exotic hadronic state once established.

Our results show that the $D^*\Sigma_c$ system with spin-parity $J^P = 1/2^-$
and isospin $I = 1/2$ can form a loosely bound state. In the isospin symmetry
case, with the momentum cutoff varied in the range $\Lambda = 1.15$ -
$1.25$~GeV, the binding energy ranges from $0.53$ to $14.38$~MeV and the
corresponding {\it rms} radius from $4.33$ to $1.14$~fm, characteristic of a
typical hadronic molecule. This state is predominantly a $^2S_{1/2}$ component,
with a probability of $86.72\%$ - $96.34\%$, and a minor admixture of the
$^4D_{1/2}$ component ($3.66\%$ - $13.28\%$).

The isospin-$3/2$ $D^*\Sigma_c$ system with $J^P = 1/2^-$ can also bind in the
isospin symmetry. For the cutoff value $\Lambda = 1.40$ - $1.90$~GeV, the
binding energy ranges from $0.22$ to $18.77$~MeV and the {\it rms} radius from
$5.11$ to $0.87$~fm. This loosely bound state is dominated by the $^2S_{1/2}$
with a probability exceeding $99\%$, while the $^4D_{1/2}$ contribution is
below $1\%$. The triply charged state $D^{*+}\Sigma_c^{++}$ with $\ket{I,I_3} =
 \ket{3/2, 3/2}$ is especially interesting, since this state cannot be formed
from three quarks in the conventional quark model. It would thus represent a
genuinely exotic baryon if observed experimentally.

The inclusion of isospin breaking leads to significant modifications in the
binding energy. For the isospin-$1/2$ case with $\Lambda = 1.22$~GeV, the
analyses show that the isospin-breaking effect provides around $17\%$
correction to the binding energy, a sizable effect given the shallow nature of
the molecular state.

For the isospin-$3/2$ case, the states $\ket{I,I_3} = \ket{3/2, 1/2}$ and
$\ket{3/2, 3/2}$ are degenerate under isospin symmetry. This degeneration is
dissolved once isospin breaking is introduced, and the resulting splitting
provides a measure of the strength of isospin-breaking effects. As in the
isospin-$1/2$ case, the binding of the $D^*\Sigma_c$ state with $J^P = 1/2^-$
and $\ket{I,I_3} = \ket{3/2, 1/2}$ becomes weaker upon the inclusion of isospin
breaking, corresponding to exceeding $30\%$ binding energy correction for this
loosely bound molecular state (see Table~\ref{1/2DstarSigma}).

\begin{table}
 \caption{Binding solutions of the $D^{*} \Sigma_{c}$ system with $J^P = 1/2^-$. The cases I and II mean without (w/o) and with (w/) isospin breaking, respectively.
 $\Lambda$, $E_b,$ and $r_{rms}$ are respective the momentum cutoff (in units of GeV), binding energy (in units of MeV) and {\it rms} radius (in units of fm). $P_{i}(\%)$
 is the probability of the channel $i$, with $i = \ket{^2 S_{1/2}} $ or $\ket{^4 D_{1/2}}$. } \label{1/2DstarSigma}
 \begin{ruledtabular}
 \begin{tabular}{ccc ccccc}
 $\ket{I,I_3}$ & \multicolumn{2}{c}{Cases} & $\Lambda$ (GeV)& $E_b$ (MeV) & $r_{rms}$ (fm) & \multicolumn{2}{c}{$P_{i} (\%)$} \\
 \midrule
 \multirow{6}{*}{$\ket{1/2, 1/2}$} & \multicolumn{2}{c}{} & $1.16$ & $0.53$ & $4.33$ & $96.34$ & $3.66$ \\
 & \multicolumn{2}{c}{I (w/o) } & $1.22$ & $7.16$ & $1.49$ & $89.18$ & $10.82$ \\
 & \multicolumn{2}{c}{} & $1.25$ & $14.38$ & $1.14$ & $86.72$ & $13.28$ \\
 \cmidrule{2-8}
 & \multicolumn{2}{c}{} & $1.17$ & $0.03$ & $3.70$ & $95.42$ & $4.58$ \\
 & \multicolumn{2}{c}{II (w/) } & $1.22$ & $5.92$ & $1.53$ & $89.29$ & $10.71$ \\
 & \multicolumn{2}{c}{} & $1.25$ & $13.02$ & $1.15$ & $86.76$ & $13.24$ \\
 \cmidrule{1-8}
 \multirow{6}{*}{$\ket{3/2, 1/2}$} & \multicolumn{2}{c}{} & $1.40$ & $0.22$ & $5.11$ & $99.78$ & $0.22$ \\
 & \multicolumn{2}{c}{I (w/o)} & $1.72$ & $8.44$ & $1.22$ & $99.40$ & $0.60$ \\
 & \multicolumn{2}{c}{} & $1.90$ & $18.77$ & $0.87$ & $99.30$ & $0.70$ \\
 \cmidrule{2-8}
 & \multicolumn{2}{c}{} & $1.55$ & $0.15$ & $2.42$ & $99.58$ & $0.42$ \\
 & \multicolumn{2}{c}{II (w/)} & $1.72$ & $5.51$ & $1.30$ & $99.41$ & $0.59$ \\
 & \multicolumn{2}{c}{} & $1.90$ & $15.41$ & $0.90$ & $99.30$ & $0.70$ \\
 \cmidrule{1-8}
 \multirow{4}{*}{$\ket{3/2, 3/2}$} & \multicolumn{2}{c}{I (w/o)} & \multicolumn{5}{c}{Same as those of $\ket{I,I_3}=\ket{3/2, 1/2}$ in case I} \\
 \cmidrule{2-8}
 & \multicolumn{2}{c}{} & $1.60$ & $0.59$ & $2.94$ & $99.59$ & $0.41$ \\
 & \multicolumn{2}{c}{II (w/)} & $1.72$ & $3.99$ & $1.52$ & $99.43$ & $0.57$ \\
 & \multicolumn{2}{c}{} & $1.90$ & $12.96$ & $0.96$ & $99.30$ & $0.70$ \\
 \end{tabular}
 \end{ruledtabular}
\end{table}

\begin{table*}
 \caption{Binding solutions of the $D^{*} \Sigma_{c}$ system with $J^P = 3/2^-$. I and II denote the cases without (w/o) and with (w/) isospin breaking, respectively.
 $\Lambda$, $E_b$ and $r_{rms}$ are the momentum cutoff (in units of GeV), the binding energy (in units of MeV) and {\it rms} radius (in units of fm).
 $P_{i} (\%)$ is the probability of the channel $i$, with $i = \ket{^4 S_{3/2}}, \ket{^2 D_{3/2}}$ or $\ket{^4 D_{3/2}}$.} \label{3/2DstarSigma}
 \begin{ruledtabular}
 \begin{tabular}{c cc cccccc}
 $\ket{I, I_3} $ & \multicolumn{2}{c}{Cases} & $\Lambda$ (GeV) & $E_b$ (MeV) & $r_{rms}$ (fm) & \multicolumn{3}{c}{$P_{i} (\%)$} \\
 \midrule
 \multirow{6}{*}{$\ket{1/2, 1/2}$} & & & $0.91$ & $0.33$ & $4.82$ & $97.86$ & $0.49$ & $1.65$ \\
 & \multicolumn{2}{c}{I (w/o)} & $1.00$ & $7.58$ & $1.46$ & $94.37$ & $1.28$ & $4.35$ \\
 & & & $1.04$ & $13.88$& $1.18$ & $93.33$ & $1.50$ & $5.16$ \\
 \cmidrule{2-9}
 & & & $0.93$ & $0.10$ & $3.47$ & $96.98$ & $0.70$ & $2.33$ \\
 & \multicolumn{2}{c}{II (w/)} & $1.00$ & $6.35$ & $1.49$ & $94.39$ & $1.27$ & $4.34$ \\
 & & & $1.04$ & $12.57$& $1.19$ & $93.33$ & $1.51$ & $5.16$ \\
 \cmidrule{1-9}
 \multirow{6}{*}{$\ket{3/2, 1/2}$} & & & $3.90$ & $0.82$ & $3.52$ & $98.08$ & $0.30$ & $1.62$ \\
 & \multicolumn{2}{c}{I (w/o)} & $4.70$ & $9.73$ & $1.20$ & $94.86$ & $0.74$ & $4.40$ \\
 & & & $4.90$ & $14.25$& $1.02$ & $94.11$ & 0.84 & $5.05$ \\
 \cmidrule{2-9}
 & & & $4.20$ & $0.33$ & $2.41$ & $97.14$ & $0.43$ & $2.43$ \\
 & \multicolumn{2}{c}{II (w/)} & $4.70$ & $5.12$ & $1.44$ & $95.40$ & $0.67$ & $3.93$ \\
 & & & $4.90$ & $11.06$& $1.06$ & $94.24$ & $0.82$ & $4.94$ \\
 \cmidrule{1-9}
 \multirow{4}{*}{$\ket{3/2, 3/2}$} & \multicolumn{2}{c}{I (w/o)} & \multicolumn{6}{c}{Same as those of $\ket{I, I_3} =\ket{3/2, 1/2}$ in case I} \\
 \cmidrule{2-9}
 & & & $4.30$ & $0.54$ & $3.13$ & $97.47$ & $0.38$ & $2.15$ \\
 & \multicolumn{2}{c}{II (w/)} & $4.70$ & $5.12$ & $1.44$ & $95.40$ & $0.67$ & $3.93$ \\
 & & & $4.90$ & $8.98$ & $1.16$ & $94.54$ & $0.78$ & $4.68$ \\
 \end{tabular}
 \end{ruledtabular}
\end{table*}

In the $J^P = 3/2^-$ sector, the isospin-$1/2$ $D^*\Sigma_c$ system forms a
weakly bound state in the isospin symmetry, with binding energy $0.33$ -
$13.88$~MeV and rms radius $4.82$ - $1.18$~fm for $\Lambda = 0.91$ -
$1.04$~GeV. This state is predominantly $^4S_{3/2}$, with small admixtures from
$^2D_{3/2}$ and $^4D_{3/2}$. No bound state is found for the isospin-$3/2$
configuration within the physically relevant cutoff region. When isospin
breaking is included, the binding of the isospin-$1/2$ state weakens, as shown
in Table~\ref{3/2DstarSigma}.

\subsection{The $D^{*} \Sigma_{c}^{*}$ systems }\label{DstarSigmastar}
The $D^*\Sigma_c^*$ system can form ground states with spin-parity $J^P = 1/2^-$, $3/2^-$, and $5/2^-$. In addition to the dominant $S$-wave configurations, we also include higher partial waves that can couple via the tensor force. Specifically, we consider the partial waves ${}^2 S_{1/2}$, ${}^4 D_{1/2}$, ${}^6 D_{1/2}$ for $J^P = 1/2^-$,
${}^4 S_{3/2}, {}^2 D_{3/2}, {}^4 D_{3/2}, {}^6 D_{3/2}$ for $J^P = 3/2^-$, and
${}^6 S_{5/2}, {}^2 D_{5/2}, {}^4 D_{5/2}, {}^6 D_{5/2}$ for $J^P = 5/2^-$, see Table~\ref{tab: SDstates}.
We present the numerical results in Tables~\ref{DstarSigmastar_1half}, \ref{DstarSigmastar_3half} and \ref{DstarSigmastar_5half} for
$J^P = 1/2^-$, $3/2^-$ and $5/2^-$, respectively.

The numerical results demonstrate that the $D^*\Sigma_c^*$ with $J^P = 1/2^-$
and $I = 1/2$ can form a hadronic molecular state. Its binding energy and {\it
 rms} radius are respective $0.26 - 14.34$~MeV and $5.32 - 1.20$~fm for the
momentum cutoff $\Lambda=1.09 - 1.19$~GeV in the isospin symmetry limit. It
dominantly lies in state $\ket{{}^2S_{1/2}}$ ($95.90\%-80.54\%$), and small
contributions stem from ${}^4D_{1/2}$ ($1.95\%-8.23\%$) and ${}^4D_{1/2}$
($2.16\%-11.22\%$). The situation for the isospin-$3/2$ case is similar. Under
isospin symmetry, the binding energy and {\it rms} radius for the $\ket{I, I_3}
 = \ket{3/2, 1/2}$ state are $0.36 -14.81$~MeV and $4.53 - 1.98$~fm
respectively, when the momentum cutoff is set to $\Lambda=1.20 - 1.55$~GeV. The
contribution of the dominant channel ${}^2S_{1/2}$ is around $90\%$. With
isospin symmetry, the states $\ket{I, I_3} = \ket{3/2, 1/2}$ and $\ket{3/2,
 3/2}$ are degenerate.

Once the isospin breaking is included, the {\it rms} radius of $D^*\Sigma_c^*$
with isospin-$1/2$ increases from $1.56$~fm to $1.59$~fm when $\Lambda =
 1.16$~GeV. Accordingly, the binding energy decreases from $7.12$~MeV to
$5.82$~MeV, which indicates that the isospin breaking leads to a relative
correction of $18.26\%$ to the binding energy. The effect is even more
pronounced for the isospin-$3/2$ states. For example, in the $\ket{I, I_3} =
 \ket{3/2, 1/2}$ channel with $\Lambda = 1.45$~GeV, the binding energy drops
from $8.45$~MeV to $5.41$~MeV upon the inclusion of isospin breaking, yielding
a relative change of approximately $36\%$ to the binding energy.

This enhanced sensitivity can be understood from the effective charge product
$N_{\rm eff}$ associated with each state. For the isospin-$1/2$ configuration,
the repulsive Coulomb interaction corresponds to $N_{\rm eff} = 1/3$, whereas
for the $\ket{I, I_3} =\ket{3/2, 3/2}$ state $D^{*+} \Sigma_{c}^{*++}$, one has
$N_{\rm eff} = 2$. The stronger Coulomb repulsion in the latter case leads to a
more significant impact on the binding energy. Among all sources of isospin
breaking considered, we find that the electromagnetic interaction plays the
dominant role.

For the $J^P = 3/2^-$ channel with isospin $I = 1/2$, a weakly bound state is
also found for cutoff values around $1.10$~GeV. Specifically, with isospin
symmetry, the {\it rms} radius and binding energy are $1.50$~fm and $7.62$~MeV
respectively for $\Lambda = 1.14$~GeV, and the results have a mild dependence
on the momentum cutoff. This state thus represents a good hadronic molecular
candidate. In contrast, for the isospin-$3/2$ configuration with the same $J^P
 = 3/2^-$, no bound solution is obtained unless the cutoff exceeds $1.80$~GeV.
As in the $J^P = 1/2^-$ case, isospin breaking weakens the binding of
$D^*\Sigma_c^*$ with $J^P = 3/2^-$ and isospin-$1/2$. At $\Lambda = 1.14$~GeV,
the binding energy decreases from $7.62$~MeV to $6.29$~MeV upon the inclusion
of isospin breaking, corresponding to a relative correction of $17.45\%$.

For the state with $J^P = 5/2^-$ and isospin-$1/2$, a loosely bound state is
obtained when $\Lambda\sim1.0$~GeV. Within isospin symmetry, its {\it rms}
radius and binding energy are $4.94-1.11$~fm and $0.28-15.79$~MeV respectively
when $\Lambda=0.85-1.00$~GeV. It mainly lies in the partial wave ${}^6S_{5/2}$,
with a probability of $98.22\%-94.54\%$. Minor contributions arise from the
${}^2D_{5/2}$ and $^4D_{5/2}$ (less than $1\%$), and small contribution stems
from the partial wave ${}^6D_{5/2}$ ($1.45\% - 4.47\%$). In contrast, no bound
solution is found for the isospin-$3/2$ configuration unless $\Lambda$ is
increased to as large as $4.0$~GeV. When isospin breaking is included, the
binding of the isospin-$1/2$ $D^*\Sigma_c^*$ state weakens. For instance, at a
fixed cutoff $\Lambda = 0.95$~GeV, the binding energy decreases from $8.00$~MeV
to $6.65$~MeV upon the inclusion of isospin breaking, corresponding to a
relative correction of $16.88\%$. In contrast, the probabilities of the
individual partial waves remain largely unaffected by isospin breaking.

\begin{table*}
 \caption{Binding solutions of the $D^{*} \Sigma_{c}^{*}$ system with $J^P = 1/2^-$. I and II denote the cases without (w/o) and with (w/) isospin breaking, respectively.
 $\Lambda$, $E_b$ and $r_{rms}$ are the respective momentum cutoff (in units of GeV), binding energy (in units of MeV) and {\it rms} radius (in units of fm).
 $P_{i} (\%)$ is the probability of the channel $i$, with $i = \ket{^2 S_{1/2}}$, $\ket{^4 D_{1/2}}$ or $\ket{^6 D_{1/2}}$.} \label{DstarSigmastar_1half}
 \begin{ruledtabular}
 \begin{tabular}{cccccccc}
 $\ket{I, I_3}$ & Cases& $\Lambda$ (GeV) & $E_b$ (MeV) & $r_{rms}$ (fm) & \multicolumn{3}{c}{$P_{i} (\%)$}\\
 \midrule
 \multirow{6}{*}{$\ket{1/2, 1/2}$} & \multirow{3}{*}{I (w/o)} & $1.09$ & $0.26$ & $5.32$ & $95.90$ & $1.95$ & $2.14$ \\
 & & $1.16$ & $7.12$ & $1.56$ & $84.10$ & $6.92$ & $8.98$ \\
 & & $1.19$ & $14.34$ & $1.20$ & $80.54$ & $8.23$ & $11.22$ \\
 \cmidrule{2-8}
 & \multirow{3}{*}{II (w/)} & $1.12$ & $0.67$ & $2.88$ & $90.93$ & $4.16$ & $4.91$ \\
 & & $1.16$ & $5.82$ & $1.59$ & $84.24$ & $6.87$ & $8.89$ \\
 & & $1.19$ & $12.94$ & $1.21$ & $80.60$ & $8.21$ & $11.19$ \\
 \cmidrule{1-8}
 \multirow{6}{*}{$\ket{ 3/2, 1/2}$} & \multirow{3}{*}{I (w/o)}
 & $1.20$ & $0.36$ & $4.53$ & $99.56$& $0.26$ & $0.17$ \\
 & & $1.45$ & $8.45$ & $1.24$ & $99.01$ & $0.61$ & $0.38$ \\
 & & $1.55$ & $14.81$ & $0.98$ & $98.92$ & $0.67$ & $0.42$ \\
 \cmidrule{2-8}
 & \multirow{3}{*}{II (w/)} & $1.32$ & $0.28$ & $2.24$ & $99.24$ & $0.46$ & $0.30$ \\
 & & $1.45$ & $5.41$ & $1.30$ & $99.02$ & $0.60$ & $0.38$ \\
 & & $1.55$ & $11.51$ & $1.01$ & $98.92$ & $0.67$ & $0.42$ \\
 \cmidrule{1-8}
 \multirow{4}{*}{$\ket{3/2, 3/2}$} & I (w/o) & \multicolumn{6}{c}{Same as those of $\ket{I, I_3} = \ket{3/2, 1/2}$ in case I} \\
 \cmidrule{2-8}
 & \multirow{3}{*}{II (w/)} & $1.34$ & $0.48$ & $3.10$ & $99.33$ & $0.41$ & $0.27$ \\
 & & $1.45$ & $4.25$ & $1.49$ & $99.04$ & $0.59$ & $0.37$ \\
 & & $1.55$ & $9.79$ & $1.09$ & $98.92$ & $0.67$ & $0.42$ \\
 \end{tabular}
 \end{ruledtabular}
\end{table*}

\begin{table*}
 \caption{Binding solutions of the $D^{*} \Sigma_{c}^{*}$ system with $J^P = 3/2^-$. I and II denote the cases without (w/o) and with (w/) isospin breaking, respectively.
 $\Lambda$, $E_b$, and $r_{rms}$ are the momentum cutoff (in units of GeV), binding energy (in units of MeV) and {\it rms} radius (in units of fm). $P_{i} (\%)$ is
 the probability of the channel $i$, with $i = \ket{^4 S_{3/2}}, \ket{^2 D_{3/2}}, \ket{^4 D_{3/2}}$ or $\ket{^6 D_{3/2}}$.} \label{DstarSigmastar_3half}
 \begin{ruledtabular}
 \begin{tabular}{cc ccccccc}
 $\ket{I, I_3}$& Cases& $\Lambda$ (GeV) & $E_b$ (MeV) & $r_{rms}$ (fm) & \multicolumn{4}{c}{$P_{i} (\%)$}\\
 \midrule
 \multirow{5}{*}{$\ket{1/2, 1/2}$} & \multirow{2}{*}{I (w/o)} & $1.14$ & $7.62$ & $1.50$ & $91.46$ & $2.93$ & $4.87$ & $0.74$ \\
 & & $1.18$ & $15.84$ & $1.14$ & $89.71$ & $3.53$ & $5.84$ & $0.91$ \\
 \cmidrule{2-9}
 & \multirow{3}{*}{II (w/)} & $1.08$ & $0.09$ & $3.44$ & $95.67$ & $1.48$ & $2.49$ & $0.35$ \\
 & & $1.14$ & $6.29$ & $1.52$ & $91.52$ & $2.91$ & $4.84$ & $0.73$ \\
 & & $1.18$ & $14.41$ & $1.15$ & $89.73$ & $3.53$ & $5.83$ & $0.91$ \\
 \cmidrule{1-9}
 \multirow{6}{*}{$\ket{3/2, 1/2}$} & \multirow{3}{*}{I (w/o)} & $1.80$ & $0.19$ & $5.33$ & $99.38$ & $0.17$ & $0.41$ & $0.05$ \\
 & & $2.30$ & $9.02$ & $1.20$ & $97.70$ & $0.61$ & $1.52$ & $0.17$ \\
 & & $2.50$ & $16.96$ & $0.92$ & $97.24$ & $0.73$ & $1.83$ & $0.20$ \\
 \cmidrule{2-9}
 & \multirow{3}{*}{II (w/)} & $2.05$ & $0.17$ & $2.31$ & $98.53$ & $0.39$ & $0.97$ & $0.11$ \\
 & & $2.30$ & $5.81$ & $1.27$ & $97.75$ & $0.59$ & $1.48$ & $0.17$ \\
 & & $2.50$ & $13.40$ & $0.95$ & $97.26$ & $0.72$ & $1.81$ & $0.20$ \\
 \cmidrule{1-9}
 \multirow{6}{*}{$\ket{3/2, 3/2}$} & I (w/o) & \multicolumn{4}{c}{Same as those of $\ket{I,I_3} = \ket{3/2, 3/2}$ in case I} \\
 \cmidrule{2-9}
 & II (w/) & $2.10$ & $0.47$ & $3.15$ & $98.71$ & $0.34$ & $0.85$ & $0.10$ \\
 & & $2.30$ & $4.38$ & $1.48$ & $97.89$ & $0.56$ & $1.40$ & $0.16$ \\
 & & $2.50$ & $11.18$ & $1.03$ & $97.33$ & $0.70$ & $1.77$ & $0.20$ \\
 \midrule
 \addlinespace
 \end{tabular}
 \end{ruledtabular}
\end{table*}

\begin{table*}
 \caption{Binding solutions of the $D^{*} \Sigma_{c}^{*}$ system with $J^P = 5/2^-$. I and II denote cases without (w/o) and with (w/) isospin breaking, respectively.
 $\Lambda$, $E_b$ and $r_{rms}$ are the respective momentum cutoff (in units of GeV), binding energy (in units of MeV) and {\it rms} radius (in units of fm).
 $P_{i}$ is the probability of the channel $i$, with $i = \ket{^6 S_{5/2}}, \ket{^2 D_{5/2}}, \ket{^4 D_{5/2}}$ or $\ket{^6 D_{5/2}}$.} \label{DstarSigmastar_5half}
 \begin{ruledtabular}
 \begin{tabular}{cc ccccccc}
 $\ket{I, I_3}$& Cases& $\Lambda$ (GeV) & $E_b$ (MeV) & $r_{rms} (fm)$ & \multicolumn{4}{c}{$P_{i} (\%)$}\\
 \midrule
 \multirow{6}{*}{$\ket{1/2, 1/2}$} & \multirow{3}{*}{I (w/o)} & $0.85$ & $0.28$ & $4.94$ & $98.22$ & $0.25$ & $0.07$ & $1.45$ \\
 & & $0.95$ & $8.00$ & $1.41$ & $95.45$ & $0.65$ & $0.18$ & $3.72$ \\
 & & $1.00$ & $15.79$ & $1.11$ & $94.54$ & $0.78$ & $0.21$ & $4.47$ \\
 \cmidrule{2-9}
 & \multirow{3}{*}{II (w/)} & $0.88$ & $0.32$ & $2.99$ & $97.18$ & $0.40$ & $0.11$ & $2.31$ \\
 & & $0.95$ & $6.65$ & $1.43$ & $95.46$ & $0.65$ & $0.18$ & $3.72$ \\
 & & $1.00$ & $14.35$ & $1.12$ & $94.53$ & $0.78$ & $0.21$ & $4.47$ \\
 \cmidrule{1-9}
 \multirow{6}{*}{$\ket{3/2, 1/2}$} & \multirow{3}{*}{I (w/o)} & $3.60$ & $0.51$ & $4.23$ & $97.93$ & $0.11$ & $0.06$ & $1.90$ \\
 & & $4.30$ & $8.22$ & $1.30$ & $93.41$ & $0.26$ & $0.20$ & $6.12$ \\
 & & $4.60$ & $16.00$ & $0.98$ & $91.58$ & $0.30$ & $0.26$ & $7.85$ \\
 \cmidrule{2-9}
 & \multirow{3}{*}{II (w/)} & $4.00$ & $0.78$ & $2.12$ & $95.67$ & $0.19$ & $0.14$ & $4.01$ \\
 & & $4.30$ & $5.18$ & $1.37$ & $93.64$ & $0.26$ & $0.20$ & $5.91$ \\
 & & $4.60$ & $12.63$ & $1.02$ & $91.74$ & $0.30$ & $0.26$ & $7.71$ \\
 \cmidrule{1-9}
 \multirow{4}{*}{$\ket{3/2, 3/2}$} & I (w/o) & \multicolumn{4}{c}{Same as those of $\ket{I, I_3} = \ket{3/2, 1/2}$ in case I} \\
 \cmidrule{2-9}
 & \multirow{3}{*}{II (w/)} & $4.05$ & $0.88$ & $2.73$ & $96.23$ & $0.17$ & $0.12$ & $3.48$ \\
 & & $4.30$ & $4.05$ & $1.59$ & $94.19$ & $0.24$ & $0.18$ & $5.40$ \\
 & & $4.60$ & $10.78$ & $1.09$ & $92.09$ & $0.29$ & $0.25$ & $7.37$ \\
 \end{tabular}
 \end{ruledtabular}
\end{table*}

\subsection{The Isospin-breaking Angle}
Under isospin symmetry, all states within a given isospin multiplet $\ket{I,
 I_3}$ are degenerate. Once isospin breaking is introduced, this degeneracy is
dissolved, and states with different total isospin $I$ but the same third
component $I_3$ can mix. Consequently, total isospin $I$ is no longer a good
quantum number, although $I_3$ remains conserved. In this subsection, we
introduce a mixing angle that quantifies the degree of isospin mixing induced
by isospin breaking. This angle also serves as a measure of the strength of
isospin-breaking effects.

To compute the mixing angle, we consider the two isospin states $\ket{I, I_3} =
 \ket{1/2, 1/2}$ and $\ket{3/2, 1/2}$ of the $D^{(*)}\Sigma_c^{(*)}$ system.
With these two bases, the off diagonal matrix elements of the Hamiltonian are
nonzero, leading to the mixing of these two isospin states. The mixing angle
$\theta$ is defined as
\begin{align}
 \ket{D^{(*)}\Sigma_c^{(*)}} \equiv & \cos{\theta}\ket{D^{(*)}\Sigma_c^{(*)}}^{(I, I_3) = (1/2, 1/2)} \nonumber \\
 & + \sin{\theta}\ket{D^{(*)}\Sigma_c^{(*)}}^{(I, I_3) = (3/2, 1/2)},
\end{align}
from which
\begin{align}
 \theta = \frac{180^\circ}{\pi} \arccos(\sqrt{P_{I=1/2}}),
\end{align}
with $P_{I=1/2}$ the probability of the isospin-$1/2$ state found in the mixed state.
When $\theta = 0^\circ$ ($90^\circ$), the state is purely isospin-$1/2$ ($3/2$), corresponding to $P_{I=1/2} = 100\%$ ($P_{I=3/2} = 100\%$). In contrast, $\theta = 45^\circ$ represents maximal isospin mixing, where the state has equal probability of being isospin-$1/2$ and isospin-$3/2$.
In Table~\ref{tab:Iso-theta}, we present the numerical results of the isospin-breaking angle $\theta$ of the $D^{(*)} \Sigma_c^{(*)}$ systems.
\begin{table}[h]
 \caption{\label{tab:Iso-theta}
 The isospin-breaking angle $\theta$ of the $D^{(*)} \Sigma_{c}^{(*)}$ systems. The unit of the cutoff $\Lambda$ is GeV.
 ``$\cdots$'' means no such states at the $S$-wave level.}
 \begin{ruledtabular}
 \begin{tabular}{lcc ccc ccc ccc}
 & \multicolumn{2}{c}{$\rm D \Sigma_{c}$} &
 & \multicolumn{2}{c}{$\rm D \Sigma_{c}^{*}$} &
 & \multicolumn{2}{c}{$\rm D^{*} \Sigma_{c}$} &
 & \multicolumn{2}{c}{$\rm D^{*} \Sigma_{c}^{*}$} \\
 $J^{P}$ & $\Lambda$ & $\theta$ & & $\Lambda$ & $\theta$ & & $\Lambda$ & $\theta$ & & $\Lambda$ & $\theta$ \\
 \hline
 \addlinespace
 & 1.18 & $17.2^{\circ}$ &
 & &&
 & 1.18 & $17.4^{\circ}$ &
 & 1.12 & $20.3^{\circ}$ \\
 $\frac{1}{2}^{-}$
 & 1.25 & $7.2^{\circ}$ &
 & $\cdots$ & $\cdots$ &
 & 1.22 & $7.6^{\circ}$ &
 & 1.16 & $9.3^{\circ}$ \\
 & 1.30 & $4.7^{\circ}$ &
 & &&
 & 1.25 & $4.6^{\circ}$ &
 & 1.19 & $5.5^{\circ}$ \\
 \hline
 \addlinespace
 & &&
 & 1.18 & $17.1^{\circ}$ &
 & 0.93 & $17.9^{\circ}$ &
 & 1.08 & $19.5^{\circ}$ \\
 $\frac{3}{2}^{-}$
 & $\cdots$ & $\cdots$ &
 & 1.25 & $7.3^{\circ}$ &
 & 1.00 & $4.8^{\circ}$ &
 & 1.14 & $6.3^{\circ}$ \\
 & &&
 & 1.30 & $4.8^{\circ}$ &
 & 1.04 & $3.0^{\circ}$ &
 & 1.18 & $3.7^{\circ}$ \\
 \hline
 \addlinespace
 & &&
 & &&
 & &&
 & 0.88 & $16.2^{\circ}$ \\
 $\frac{5}{2}^{-}$
 & $\cdots$ & $\cdots$ &
 & $\cdots$ & $\cdots$ &
 & $\cdots$ & $\cdots$ &
 & 0.95 & $4.9^{\circ}$ \\
 & &&
 & &&
 & &&
 & 1.00 & $2.9^{\circ}$ \\
 \end{tabular}
 \end{ruledtabular}
\end{table}

Our results show that the isospin mixing angle lies approximately in the range
$\theta=3^\circ$-$20^\circ$. Notably, $\theta$ increases as the binding energy
decreases: the more weakly bound the system, the larger the mixing angle. This
implies that for hadronic molecular states, which are typically characterized
by small binding energies, isospin-breaking effects are amplified. Such effects
must therefore be incorporated in high-precision calculations to match the
demands of rapidly developing experiments.

\section{Conclusion} \label{summary}

In this work, we investigate the role of explicit isospin breaking in
deuteron-like molecular states of the $D^{(*)}\Sigma_c^{(*)}$ systems within
the framework of the OBE model. The isospin-breaking effects considered include
the Coulomb interaction between charged hadrons, the threshold differences
among components within the same isospin multiplet, and the mass splittings
between charged and neutral exchanged light mesons--specifically, the
pseudoscalar mesons $\pi^{\pm}$ and $\pi^0$, as well as the vector mesons
$\rho^{\pm}$ and $\rho^0$.

Our results show that all four $D^{(*)}\Sigma_c^{(*)}$ systems with isospin $I
 = 1/2$ form well-bound states, with binding energies of several MeV for
physically reasonable cutoff values. In contrast, for the isospin-$3/2$
configurations, the contributions from $\rho$ and $\omega$ exchange largely
cancel each other, resulting in a weak net interaction. Consequently, only the
$D^*\Sigma_c$ and $D^*\Sigma_c^{*}$ systems with $J^P = 1/2^-$ are able to form
weakly bound states within the reasonable momentum cutoff values. For these
molecular candidates, we have explicitly included isospin-breaking effects: the
Coulomb interaction between charged constituents, threshold differences among
the $D^{(*)}\Sigma_c^{(*)}$ components within the same isospin multiplet, and
the mass splittings between charged and neutral pions and rho mesons in the
exchange potentials. Owing to the shallow nature of the bound states, isospin
breaking is found to induce a correction of $10\%$–$30\%$ to the binding
energy. This sizable effect underscores the necessity of incorporating isospin
breaking in high-precision theoretical calculations aimed at keeping pace with
rapid experimental progress.

We also compute the mixing angle between different isospin states, i.e., $I =
 1/2$ and $I = 3/2$ configurations, since the total isospin $I$ is no longer
conserved while its third component $I_3$ remains a good quantum number. We
find that for very loosely bound states with small binding energies, this
mixing angle can reach values as large as $18^\circ$-$20^\circ$. However, the
mixing angle rapidly decreases as the binding energy increases by finely tuning
the momentum cutoff.

As experimental data on exotic hadronic states continue to accumulate,
high-precision theoretical calculations that incorporate isospin-breaking
effects are urgently needed. We hope that the present study of isospin breaking
in hadronic molecular states will provide valuable guidance for future
experimental searches for deuteron-like hadron molecules.

\begin{acknowledgments}
 This work is supported by the Special Funds for Theoretical Physics under the National Natural Science Foundation of China (Grant No. 12547105), the National Natural Science Foundation of China (Grant No. 12575153).
\end{acknowledgments}

\appendix

\section{\label{app:1} Details of OBE potential model}

\subsection{The Meson Fields}
In the OBE potential model, some of physical variables can be simplified.
Expand $\hat{\xi}^{\mu}$ in the axial and vector currents upto leading order,
we obtain
\begin{eqnarray}
 \hat{\xi} \sim 1+ {\rm i} \mathbb{P} /f_{\pi } , \quad \hat{\xi }^{\dagger } \sim 1 - {\rm i} \mathbb{P} / f_{\pi }, \\
 \hat{X}^{\rho} \sim {\rm i} \partial^{\rho} \mathbb{P} / f_{\pi }, \quad \hat{V}^{\rho} \sim \mathbb{P} \partial^{\rho} \mathbb{P} / f_{\pi }^2,
\end{eqnarray}
from which it is clear that there is no contribution from $\hat{V}^{\rho}$ at leading order. Similarly, for $\hat{F}^{\alpha \beta}$
we can omit the two-boson-exchange term $[\mathbb{V} ^{\prime \alpha } , \mathbb{V}^{\prime \beta } ]$, but keep $\hat{F}^{\alpha \beta} = \partial^{\alpha } \mathbb{V}^{\prime \beta} - \partial ^{\beta} \mathbb{V}^{\prime \alpha }$.

\subsection{The Spin Current Operators}
The Feynmann rules of OBE potential model are
\begin{eqnarray}
 &&\begin{cases}
 \bra{0} {\rm M} \ket{M} \equiv \sqrt{m_{[M]}}, \\
 \bra{M} {\rm M} \ket{0} \equiv \sqrt{m_{[M]}},
 \end{cases} \\
 &&\begin{cases}
 \bra{0} {\rm M^{* \mu}} \ket{M^*} \equiv \sqrt{m_{[M^*]}} \varepsilon^{\mu}(\lambda), \\
 \bra{M^*} {\rm M^{* \mu}} \ket{0} \equiv \sqrt{m_{[M^*]}} \bar{\varepsilon}^{\mu}(\lambda),
 \end{cases} \\
 && \begin{cases}
 \bra{0} {\rm B} \ket{\cal{B}} \equiv \sqrt{2m_{[\cal{B}]}} u_{\rm{B}}(p,s), \\
 \bra{\cal{B}} {\rm B} \ket{0} \equiv \sqrt{2m_{[\cal{B}]}} \bar{u}_{\rm{B}}(p,s),
 \end{cases} \\
 && 	\begin{cases}
 \bra{0} {\rm B}^{* \mu} \ket{\cal{B}^{*}} \equiv \sqrt{2m_{[\cal{B}^*]}} U^{\mu}_{\rm{B}^{*}}(p,m_j), \\
 \bra{\cal{B}^{*}} {\rm B}^{* \mu} \ket{0} \equiv \sqrt{2m_{[\cal{B}^*]}} \bar{U}^{\mu}_{\rm{B}^{*}}(p,m_j),
 \end{cases}
\end{eqnarray}
where the polarization vectors, spinors and vector-spinors are defined respectively as follows:
\begin{eqnarray}
 && \epsilon ^{\mu}(+,0,-) \equiv
 \frac{1}{\sqrt{2}} \begin{bmatrix} 0\\-1\\-\rm{i}\\0 \end{bmatrix},
 \begin{bmatrix} 0\\0\\0\\1 \end{bmatrix},
 \frac{1}{\sqrt{2}} \begin{bmatrix} 0\\1\\-\rm{i}\\0 \end{bmatrix}, \\
 && \begin{cases}
 u_{\rm{B}}(p,s) = \sqrt{\frac{E_{\rm{B}} +m_{\rm{B}}}{2m_{\rm{B}}}}
 \begin{pmatrix} \chi_{s} \\
 \tfrac{\vec{\sigma} \cdotp \vec{p}}{E_{\rm{B}} + m_{\rm{B}}} \chi_{s}
 \end{pmatrix}, \\
 \bar{u}_{\rm{B}}(p^{\prime},s^{\prime}) = \sqrt{\frac{E^{\prime}_{\rm{B}} +m_{\rm{B}}}{2m_{\rm{B}}}}
 \begin{pmatrix} \chi^{\dagger}_{s^{\prime}},
 & \tfrac{\vec{\sigma} \cdotp \vec{p}^{\prime}}{E^{\prime}_{\rm{B}} + m_{\rm{B}}} \chi^{\dagger}_{s^{\prime}}
 \end{pmatrix},
 \end{cases} \\
 && \begin{cases}
 U_{\rm{B}^{*}}^{\mu}(\vec{p}, \phi) =
 \begin{pmatrix}
 \chi ^{\mu }(\phi) \\
 \tfrac{\vec{\sigma } \cdotp \vec{p}}{2m_{\rm{B}^{*}}} \chi ^{\mu}(\phi)
 \end{pmatrix}, \\
 \bar{U}_{\rm{B}^{*}}^{\mu}(\vec{p}^{\prime}, \phi^{\prime}) =
 \begin{pmatrix}
 \chi ^{\dagger \mu}(\phi^{\prime}), &
 \tfrac{\vec{\sigma} \cdotp \vec{p}^{\prime}}{2m_{\rm{B}^{*}}} \chi ^{\dagger \mu}(\phi^{\prime})
 \end{pmatrix}.
 \end{cases}
\end{eqnarray}
The single spinors and single vector-spinors read respectively as
\begin{align}
 & \chi_{1/2} =\begin{pmatrix} 1\\0 \end{pmatrix}, \quad
 \chi _{-1/2} =\begin{pmatrix} 0\\1 \end{pmatrix},\\
 & \chi^{\mu }(\phi) \equiv \sum _{m_{1}, m_{2}} {\rm CG} \left[ 1m_{1} ,\tfrac{1}{2} m_{2} ,\tfrac{3}{2} \phi \right] \epsilon ^{\mu }( m_{1}) \chi ( m_{2}),
\end{align}
along with $\rm{CG}[\ldots]$ representing the Clebsch-Gordan coefficient.

In addition, we also introduce some simplified notation, which absorbs the mass
parameters in the normalization into the Breit's formula,
\begin{eqnarray}
 V({\bf q}) = - \mathcal{M} (12 \to 34),
\end{eqnarray}
with
\begin{eqnarray}
 && \begin{cases}
 \bra{0} {\rm M} \ket{M} = 1/ \sqrt{2}, \quad
 \bra{0} {\rm M^{* \mu}} \ket{M^*} = \varepsilon^{\mu}_{\lambda} / \sqrt{2}, \\
 \bra{M} {\rm M} \ket{0} = 1/ \sqrt{2}, \quad
 \bra{M^*} {\rm M^{* \mu}} \ket{0} = \bar{\varepsilon}^{\mu}_{\lambda} / \sqrt{2} ,
 \end{cases} \\
 && \begin{cases}
 \bra{0} {\rm B} \ket{\cal{B}} = u_{\rm{B}}, \quad
 \bra{0} {\rm B}^{* \mu} \ket{\cal{B}^{*}} = U^{\mu}_{\rm{B}^{*}}, \\
 \bra{\cal{B}} {\rm B} \ket{0} = \bar{u}_{\rm{B}}, \quad
 \bra{\cal{B}^{*}} {\rm B}^{* \mu} \ket{0} = \bar{U}^{\mu}_{\rm{B}^{*}}.
 \end{cases}
\end{eqnarray}

\subsection{The Potentials in Momentum Space }

Recall the expression of $n$-pole form factor, we can rewrite its form as
\begin{align}
    &F_n(\lambda, u, {\bf q}) = \left(\frac{\lambda^{2} - u^{2}}{\lambda^{2} + {\bf q}^{2}} \right)^{n}, \\
    &\lambda^2 = \Lambda^2 - q_0^2, \quad
    u^2 = m^2 - q_0^2.
\end{align}
Separating the operators from the OBE potential for the heavy-flavor hadrons,
we obtain three types of momentum currents,
\begin{eqnarray}
 && V_{0}({\bf q}) \sim \frac{1}{{\bf q}^{2} + u^{2}}, \quad
 V_{2}({\bf q}) \sim \frac{q^i q^j}{{\bf q}^{2} + u^{2}}, \nonumber \\
 && V_{\rm S}({\bf q}) \sim \frac{{\bf q}^{2}}{{\bf q}^{2} + u^{2}} = \delta^{ij} \frac{q^i q^j}{{\bf q}^{2} + u^{2}}.
\end{eqnarray}
In the above, ${\bf q}$ is the momentum of the exchanged light meson. Combining the monopole form factor $F_1(\lambda, u, {\bf q})$ and making the
Fourier transform (taking away the general factor $1 / 4 \pi$), we have
\begin{eqnarray}
 && V_{0}({\bf q}) \to V_{0}(r), \quad
 V_{\rm S}({\bf q}) \to V_{\rm S}(r), \nonumber \\
 && V_{2}({\bf q}) \to V_{2}(r) = \frac{1}{3} \delta^{ij} V_{\rm S}(r) + \frac{1}{3} \hat{T}^{ij} V_{\rm T}(r),
\end{eqnarray}
where $\hat{T}^{ij} \equiv 3 \hat{r}^{i} \hat{r}^{j} - \delta^{ij}$ and $\hat{r}$ is a unit direction vector.
The basic functions can be written explicitly as
\begin{align}
 V_{0}(r) & = uY(ur) -\lambda Y(\lambda r) -\kappa _{0} e^{-\lambda r}, \\
 V_{\rm{S}}(r) & = u^{3} Y(ur) -\lambda^{3} Y(\lambda r) -\kappa _{2}(\lambda r - 2) Y(\lambda r), \\
 V_{\rm{T}}(r) & = u^{3} Y(ur) H(ur) -\lambda^{3} Y(\lambda r) H(\lambda r) \nonumber \\
& - \kappa_{2}(\lambda r + 1) Y(\lambda r),
\end{align}
with some definitions,
\begin{align}
 Y(x) &\equiv e^{-x} / x , & H(x) &\equiv 1 + 3/x + 3/x^2, \\
 \kappa_{m} &\equiv \beta^{2} \lambda^{m-1} / 2, &\beta^2 &\equiv \lambda^2 - u^2.
\end{align}

In derivation of the effective potentials, we perform partial wave analysis to
obtain the matrices with respect different partial waves.
Then the potential can be written in the following form,
\begin{eqnarray}
 \label{eq:V} {\cal V}_{ab}^{J}(r, \Lambda, m) &\equiv& V_{0}(r) \mathbb{C}_{ab}^{J}, \\
 \label{eq:P} {\cal P}_{ab}^{J}(r, \Lambda, m) &\equiv& \frac{1}{3} V_{\rm S}(r) \mathbb{S}_{ab}^{J} + \frac{1}{3} V_{\rm T}(r) \mathbb{T}_{ab}^{J}, \\
 \label{eq:T} {\cal T}_{ab}^{J}(r, \Lambda, m) &\equiv& \frac{2}{3} V_{\rm S}(r) \mathbb{S}_{ab}^{J} - \frac{1}{3} V_{\rm T}(r) \mathbb{T}_{ab}^{J},
\end{eqnarray}
where $\mathbb{C, S, T}$ are spin matrices. $J$ is the total angular momentum while $a,b$ are indices of different systems, i.e., we use $11,22,33$ and $44$
to refer respectively to $D \Sigma_c, D \Sigma^*_c, D^* \Sigma_c$, and $D^* \Sigma^*_c$.

\section{\label{app:3} Explicit Expressions of Spin currents, and OBE Potentials}

The specific OBE potentials of $D^{(*)} \Sigma_{c}^{(*)}$ systems are

\begin{align}
 V_{11} &= \frac{1}{4 \pi} (-G_{\rm S} V_{11}^{\rm S} + \frac{1}{2} G_{\rm V} V_{11}^{\rm V}), \\
 V_{22} &= \frac{1}{4 \pi} (-G_{\rm S} V_{22}^{\rm S} + \frac{1}{2} G_{\rm V} V_{22}^{\rm V}), \\
 V_{33} &= \frac{1}{4 \pi} [(-G_{\rm S} V_{33}^{\rm S} + \frac{1}{2} G_{\rm V} V_{33}^{\rm V}) - (G_{\rm P} V_{33}^{\rm P} + \frac{2}{3} G_{\rm T} V_{33}^{\rm T})], \\
 V_{44} &= \frac{1}{4 \pi} [(-G_{\rm S} V_{44}^{\rm S} + \frac{1}{2} G_{\rm V} V_{44}^{\rm V}) - \frac{3}{2}(G_{\rm P} V_{44}^{\rm P} + \frac{2}{3} G_{\rm T} V_{44}^{\rm T})],
\end{align}
in which the products of coupling constants are
\begin{align}
 G_{\rm S} & = g_{\rm S} l_{\rm S} , & G_{\rm V} & = \beta \beta _{\rm S} g_{\rm V}^{2}, \nonumber \\
 G_{\rm P} & = gg_{1} / f_{\pi }^{2}, & G_{\rm T} & = \lambda \lambda_{\rm S} g_{\rm V}^{2}.
\end{align}
For each type, there exist
\begin{align}
 V_{ab}^{\rm S} & \equiv {\cal V}_{ab}^{J}(m_{\sigma}), \\
 V_{ab}^{\rm V} & \equiv \frac{1}{2} {\cal V}_{ab}^{J}(m_{\omega}) + {\cal I} \cdot {\cal V}_{ab}^{J}(m_{\rho}), \\
 V_{ab}^{\rm P} & \equiv \frac{1}{6} {\cal P}_{ab}^{J}(m_{\pi}) + {\cal I} \cdot {\cal P}_{ab}^{J}(m_{\eta}), \\
 V_{ab}^{\rm T} & \equiv \frac{1}{2} {\cal T}_{ab}^{J}(m_{\omega}) + {\cal I} \cdot {\cal T}_{ab}^{J}(m_{\rho}),
\end{align}
where ${\cal I}$ is the isospin factor which are
\begin{align}
 {\cal I} =
 \begin{cases}
 1, & I_{ab} =1/2, \\
 -1/2, & I_{ab} =3/2,
 \end{cases}
\end{align}
and ${\cal V}_{ab}^{J}, {\cal P}_{ab}^{J}, {\cal T}_{ab}^{J}$ are defined as Eqs.~(\ref{eq:V}, \ref{eq:P}, \ref{eq:T}) .

The spin currents for the $D^{(*)} \Sigma_{c}^{(*)}$ systems can be written as
\begin{align}
 \hat{\mathbb C}_{11} & = \hat{\cal C} (1, \chi_{4}^{\dagger} \chi_{2} ),
 & \hat{\mathbb C}_{22}& = \hat{\cal C} (1, \vec{\chi}_{4}^{\dagger} \cdot \vec{\chi}_{2} ), \nonumber \\
 \hat{\mathbb C}_{33} & = \hat{\cal C} (\vec{\epsilon }_{3}^{*} \cdot \vec{\epsilon}_{1} , \chi_{4}^{\dagger} \chi_{2} ),
 & \hat{\mathbb C}_{44}& = \hat{\cal C} (\vec{\epsilon}_{3}^{*} \cdot \vec{\epsilon}_{1} , \vec{\chi}_{4}^{\dagger} \cdot \vec{\chi}_{2} ), \\
 \hat{\mathbb S}_{33} & =\hat{\cal S} ({\rm i}\vec{\epsilon}_{3}^{*} \times \vec{\epsilon}_{1} , \chi_{4}^{\dagger}\vec{\sigma} \chi_{2} ),
 & \hat{\mathbb S}_{44}& =\hat{\cal S} ({\rm i}\vec{\epsilon}_{3}^{*} \times \vec{\epsilon}_{1} ,-{\rm i}\vec{\chi}_{4}^{\dagger } \times \vec{\chi}_{2} ), \\
 \hat{\mathbb T}_{33} & =\hat{\cal T} ({\rm i}\vec{\epsilon}_{3}^{*} \times \vec{\epsilon}_{1} , \chi_{4}^{\dagger}\vec{\sigma} \chi_{2} ),
 & \hat{\mathbb T}_{44}& =\hat{\cal T} ({\rm i}\vec{\epsilon}_{3}^{*} \times \vec{\epsilon}_{1} , -{\rm i}\vec{\chi }_{4}^{\dagger} \times \vec{\chi}_{2} ).
\end{align}
where 
\begin{align}
  \hat{\cal C}(A,B) &\equiv AB, \quad 
  \hat{\cal S}(\vec{A},\vec{B}) \equiv \vec{A} \cdot \vec{B}, \nonumber \\
  \hat{\cal T}(\vec{A},\vec{B}) &\equiv 3(\vec{A} \cdot \vec{r}) (\vec{B} \cdot \vec{r}) - \vec{A} \cdot \vec{B}.
\end{align}
One obtains the final matrix elements by projecting on the specific initial and final spin states
\begin{align}
 {\mathbb O}_{ab} = \bra{L^{\prime}S^{\prime}J^{\prime}m^{\prime}} \hat{\mathbb O}_{ab} \ket{LSJm}.
\end{align}

The nontrivial matrices ${\mathbb S}_{ab}$ and ${\mathbb T}_{ab}$ are listed in
Table.~\ref{tab: Spin Mtrx}, while all of the ${\mathbb C}_{ab}$ matrices are
just identity matrices.

\begin{table*}[]
 \caption{\label{tab: Spin Mtrx}
 Nontrivial spin matrices of $D^{(*)} \Sigma_{c}^{(*)}$ systems with $J = \frac{1}{2}, \frac{3}{2}, \frac{5}{2}$. ``$\times$'' means no such matrices at the leading order.}

 \begin{ruledtabular}
 \begin{tabular}{cccc}

 & $J^P = \frac{1}{2}^-$ & $J^P = \frac{3}{2}^-$ & $J^P = \frac{5}{2}^-$ \\

 \addlinespace
 \hline

 ${\mathbb S}_{33}$
 & $\begin{pmatrix} 2 & \\ & -1 \end{pmatrix}$
 & $\begin{pmatrix} -1 & & \\ & 2 & \\ & & -1 \end{pmatrix}$
 & $\times$ \\
 \addlinespace
 ${\mathbb T}_{33}$
 & $\begin{pmatrix} & \sqrt{2} \\ \sqrt{2} & 2 \end{pmatrix}$
 & $\begin{pmatrix} & -1 & -2 \\ -1 & & 1 \\ -2 & 1 & \end{pmatrix}$
 & $\times$ \\
 \addlinespace
 ${\mathbb S}_{44}$
 & $\begin{pmatrix} \tfrac{5}{3} & & \\ & \tfrac{2}{3} & \\ & & -1 \end{pmatrix}$
 & $\begin{pmatrix}
 \tfrac{2}{3} & & & \\
 & \tfrac{5}{3} & & \\
 & & \tfrac{2}{3} & \\
 & & & -1
 \end{pmatrix}$
 & $\begin{pmatrix}
 -1 & & & \\
 & \tfrac{5}{3} & & \\
 & & \tfrac{2}{3} & \\
 & & & -1
 \end{pmatrix}$ \\
 \addlinespace
 ${\mathbb T}_{44}$
 & $\begin{pmatrix}
 0 & \tfrac{-7}{\sqrt{45}} & \tfrac{2}{\sqrt{5}} \\
 \tfrac{-7}{\sqrt{45}} & \tfrac{16}{15}& -\tfrac{1}{5} \\
 \tfrac{2}{\sqrt{5}} & -\tfrac{1}{5} & \tfrac{8}{5}
 \end{pmatrix}$
 & $\left(\begin{array}{ c c c c }
 0 & \tfrac{7}{\sqrt{90}} & -\tfrac{16}{15} & \tfrac{-\sqrt{14}}{10} \\
 \tfrac{7}{\sqrt{90}} & 0 & \tfrac{-7}{\sqrt{90}} & \tfrac{-2}{\sqrt{35}} \\
 -\tfrac{16}{15}& \tfrac{-7}{\sqrt{90}} & 0 & \tfrac{-1}{\sqrt{14}} \\
 \tfrac{-\sqrt{14}}{10} & \tfrac{-2}{\sqrt{35}} & \tfrac{-1}{\sqrt{14}} & \tfrac{4}{7}
\end{array}\right)$
 & $\left(\begin{array}{ c c c c }
 0 & \tfrac{2}{\sqrt{15}} & \tfrac{\sqrt{21}}{15} & \tfrac{-2\sqrt{14}}{5} \\
 \tfrac{2}{\sqrt{15}} & 0 & \tfrac{\sqrt{35}}{15} & -\sqrt{\tfrac{32}{105}} \\
 \tfrac{\sqrt{21}}{15} & \tfrac{\sqrt{35}}{15} & -\tfrac{16}{21} & \tfrac{-\sqrt{6}}{21} \\
 \tfrac{-2\sqrt{14}}{5} & -\sqrt{\tfrac{32}{105}} & \tfrac{-\sqrt{6}}{21} & -\tfrac{4}{7}
\end{array}\right)$
 \end{tabular}
 \end{ruledtabular}
\end{table*}

\bibliography{reference}

@article{Yang:2024okq,
    author = "Yang, Zi-Yan and Wang, Qian and Chen, Wei",
    title = "{Mass spectra of strange double charm pentaquarks with strangeness $S = - 1$}",
    eprint = "2405.09067",
    archivePrefix = "arXiv",
    primaryClass = "hep-ph",
    doi = "10.1103/PhysRevD.110.056022",
    journal = "Phys. Rev. D",
    volume = "110",
    number = "5",
    pages = "056022",
    year = "2024"
}

@misc{Yang:2025aer,
    author = "Yang, Zi-Yan and Chen, Wei",
    title = "{Decay and production properties of strange double charm pentaquark}",
    eprint = "2511.04026",
    archivePrefix = "arXiv",
    primaryClass = "hep-ph",
    month = "11",
    year = "2025"
}

@article{Chen:2015moa,
    author = "Chen, Hua-Xing and Chen, Wei and Liu, Xiang and Steele, T. G. and Zhu, Shi-Lin",
    title = "{Towards exotic hidden-charm pentaquarks in QCD}",
    eprint = "1507.03717",
    archivePrefix = "arXiv",
    primaryClass = "hep-ph",
    doi = "10.1103/PhysRevLett.115.172001",
    journal = "Phys. Rev. Lett.",
    volume = "115",
    number = "17",
    pages = "172001",
    year = "2015"
}

@article{Belle:2003nnu,
    author = {Choi, S. K. and others},
    collaboration = {Belle},
    title = "{Observation of a narrow charmonium-like state in exclusive $B^\pm \to K^\pm \pi^+ \pi^- J/\psi$ decays}",
    eprint = {hep-ex/0309032},
    archivePrefix = {arXiv},
    doi = {10.1103/PhysRevLett.91.262001},
    journal = {Phys. Rev. Lett.},
    volume = {91},
    pages = {262001},
    year = {2003}
}

@article{BESIII:2013ris,
    author = {Ablikim, M. and others},
    collaboration = {BESIII},
    title = "{Observation of a Charged Charmoniumlike Structure in $e^+e^- \to \pi^+\pi^- J/\psi$ at $\sqrt{s}$ = 4.26 GeV}",
    eprint = "1303.5949",
    archivePrefix = "arXiv",
    primaryClass = "hep-ex",
    doi = "10.1103/PhysRevLett.110.252001",
    journal = "Phys. Rev. Lett.",
    volume = "110",
    pages = "252001",
    year = "2013"
}

@article{Belle:2013yex,
    author = "Liu, Z. Q. and others",
    collaboration = "Belle",
    title = "{Study of $e^+e^- \to \pi^+\pi^- J/\psi$ and Observation of a Charged Charmoniumlike State at Belle}",
    eprint = "1304.0121",
    archivePrefix = "arXiv",
    primaryClass = "hep-ex",
    reportNumber = "BELLE-PREPRINT-2013-6, KEK-PREPRINT-2013-2",
    doi = "10.1103/PhysRevLett.110.252002",
    journal = "Phys. Rev. Lett.",
    volume = "110",
    pages = "252002",
    year = "2013",
    note = "[Erratum: Phys.Rev.Lett. 111, 019901 (2013)]"
}

@misc{LHCb_2022,
      title="{Exotic hadron naming convention}", 
      author={LHCb collaboration},
      year={2023},
      eprint={2206.15233},
      archivePrefix={arXiv},
      primaryClass={hep-ex},
      url={https://arxiv.org/abs/2206.15233}, 
}

@article{Chen_2016,
   title="{The hidden-charm pentaquark and tetraquark states}",
   volume={639},
   ISSN={0370-1573},
   url={http://dx.doi.org/10.1016/j.physrep.2016.05.004},
   DOI={10.1016/j.physrep.2016.05.004},
   journal={Physics Reports},
   publisher={Elsevier BV},
   author={Chen, Hua-Xing and Chen, Wei and Liu, Xiang and Zhu, Shi-Lin},
   year={2016},
   month=jun, pages={1–121}
}

@article{Hosaka:2016pey,
    author = "Hosaka, Atsushi and Iijima, Toru and Miyabayashi, Kenkichi and Sakai, Yoshihide and Yasui, Shigehiro",
    title = "{Exotic hadrons with heavy flavors: X, Y, Z, and related states}",
    eprint = "1603.09229",
    archivePrefix = "arXiv",
    primaryClass = "hep-ph",
    reportNumber = "J-PARC-TH-0046",
    doi = "10.1093/ptep/ptw045",
    journal = "PTEP",
    volume = "2016",
    number = "6",
    pages = "062C01",
    year = "2016"
}

@article{Swanson:2006st,
    author = "Swanson, Eric S.",
    title = "{The New heavy mesons: A Status report}",
    eprint = "hep-ph/0601110",
    archivePrefix = "arXiv",
    doi = "10.1016/j.physrep.2006.04.003",
    journal = "Phys. Rept.",
    volume = "429",
    pages = "243--305",
    year = "2006"
}

@article{Chen:2016spr,
    author = "Chen, Hua-Xing and Chen, Wei and Liu, Xiang and Liu, Yan-Rui and Zhu, Shi-Lin",
    title = "{A review of the open charm and open bottom systems}",
    eprint = "1609.08928",
    archivePrefix = "arXiv",
    primaryClass = "hep-ph",
    doi = "10.1088/1361-6633/aa6420",
    journal = "Rept. Prog. Phys.",
    volume = "80",
    number = "7",
    pages = "076201",
    year = "2017"
}

@article{Esposito:2016noz,
    author = "Esposito, A. and Pilloni, A. and Polosa, A. D.",
    title = "{Multiquark Resonances}",
    eprint = "1611.07920",
    archivePrefix = "arXiv",
    primaryClass = "hep-ph",
    reportNumber = "JLAB-THY-16-2301",
    doi = "10.1016/j.physrep.2016.11.002",
    journal = "Phys. Rept.",
    volume = "668",
    pages = "1--97",
    year = "2017"
}

@article{Olsen:2017bmm,
    author = "Olsen, Stephen Lars and Skwarnicki, Tomasz and Zieminska, Daria",
    title = "{Nonstandard heavy mesons and baryons: Experimental evidence}",
    eprint = "1708.04012",
    archivePrefix = "arXiv",
    primaryClass = "hep-ph",
    doi = "10.1103/RevModPhys.90.015003",
    journal = "Rev. Mod. Phys.",
    volume = "90",
    number = "1",
    pages = "015003",
    year = "2018"
}

@article{Brambilla:2019esw,
    author = "Brambilla, Nora and Eidelman, Simon and Hanhart, Christoph and Nefediev, Alexey and Shen, Cheng-Ping and Thomas, Christopher E. and Vairo, Antonio and Yuan, Chang-Zheng",
    title = "{The $XYZ$ states: experimental and theoretical status and perspectives}",
    eprint = "1907.07583",
    archivePrefix = "arXiv",
    primaryClass = "hep-ex",
    reportNumber = "TUM-EFT 125/19",
    doi = "10.1016/j.physrep.2020.05.001",
    journal = "Phys. Rept.",
    volume = "873",
    pages = "1--154",
    year = "2020"
}

@article{Liu:2019zoy,
    author = "Liu, Yan-Rui and Chen, Hua-Xing and Chen, Wei and Liu, Xiang and Zhu, Shi-Lin",
    title = "{Pentaquark and Tetraquark states}",
    eprint = "1903.11976",
    archivePrefix = "arXiv",
    primaryClass = "hep-ph",
    doi = "10.1016/j.ppnp.2019.04.003",
    journal = "Prog. Part. Nucl. Phys.",
    volume = "107",
    pages = "237--320",
    year = "2019"
}

@article{Chen:2022asf,
    author = "Chen, Hua-Xing and Chen, Wei and Liu, Xiang and Liu, Yan-Rui and Zhu, Shi-Lin",
    title = "{An updated review of the new hadron states}",
    eprint = "2204.02649",
    archivePrefix = "arXiv",
    primaryClass = "hep-ph",
    doi = "10.1088/1361-6633/aca3b6",
    journal = "Rept. Prog. Phys.",
    volume = "86",
    number = "2",
    pages = "026201",
    year = "2023"
}

@article{HXChen_2021,
   title="{Establishing the first hidden-charm pentaquark with strangeness}",
   volume={81},
   ISSN={1434-6052},
   url={http://dx.doi.org/10.1140/epjc/s10052-021-09196-4},
   DOI={10.1140/epjc/s10052-021-09196-4},
   number={5},
   journal={The European Physical Journal C},
   publisher={Springer Science and Business Media LLC},
   author={Chen, Hua-Xing and Chen, Wei and Liu, Xiang and Liu, Xiao-Hai},
   year={2021},
   month=may }

@article{LHCb:2015yax,
    author = "Aaij, Roel and others",
    collaboration = "LHCb",
    title = "{Observation of $J/\psi p$ Resonances Consistent with Pentaquark States in $\Lambda_b^0 \to J/\psi K^- p$ Decays}",
    eprint = "1507.03414",
    archivePrefix = "arXiv",
    primaryClass = "hep-ex",
    reportNumber = "CERN-PH-EP-2015-153, LHCB-PAPER-2015-029",
    doi = "10.1103/PhysRevLett.115.072001",
    journal = "Phys. Rev. Lett.",
    volume = "115",
    pages = "072001",
    year = "2015"
}

@article{LHCb:2019kea,
    author = "Aaij, Roel and others",
    collaboration = "LHCb",
    title = "{Observation of a narrow pentaquark state, $P_c(4312)^+$, and of two-peak structure of the $P_c(4450)^+$}",
    eprint = "1904.03947",
    archivePrefix = "arXiv",
    primaryClass = "hep-ex",
    reportNumber = "LHCb-PAPER-2019-014 CERN-EP-2019-058",
    doi = "10.1103/PhysRevLett.122.222001",
    journal = "Phys. Rev. Lett.",
    volume = "122",
    number = "22",
    pages = "222001",
    year = "2019"
}

@article{LHCb:2020jpq,
    author = "Aaij, Roel and others",
    collaboration = "LHCb",
    title = "{Evidence of a $J/\psi\Lambda$ structure and observation of excited $\Xi^-$ states in the $\Xi^-_b \to J/\psi\Lambda K^-$ decay}",
    eprint = "2012.10380",
    archivePrefix = "arXiv",
    primaryClass = "hep-ex",
    reportNumber = "LHCb-PAPER-2020-039, CERN-EP-2020-233",
    doi = "10.1016/j.scib.2021.02.030",
    journal = "Sci. Bull.",
    volume = "66",
    pages = "1278--1287",
    year = "2021"
}

@article{LHCb:2022ogu,
    author = "Aaij, R. and others",
    collaboration = "LHCb",
    title = "{Observation of a $J / \psi \Lambda$ Resonance Consistent with a Strange Pentaquark Candidate in $B^{-} \to J / \psi \Lambda p^{-}$ Decays}",
    eprint = "2210.10346",
    archivePrefix = "arXiv",
    primaryClass = "hep-ex",
    reportNumber = "CERN-EP-2022-198, LHCb-PAPER-2022-031",
    doi = "10.1103/PhysRevLett.131.031901",
    journal = "Phys. Rev. Lett.",
    volume = "131",
    number = "3",
    pages = "031901",
    year = "2023"
}

@article{LHCb:2021vvq,
    author = "Aaij, Roel and others",
    collaboration = "LHCb",
    title = "{Observation of an exotic narrow doubly charmed tetraquark}",
    eprint = "2109.01038",
    archivePrefix = "arXiv",
    primaryClass = "hep-ex",
    reportNumber = "CERN-EP-2021-165, LHCb-PAPER-2021-031",
    doi = "10.1038/s41567-022-01614-y",
    journal = "Nature Phys.",
    volume = "18",
    number = "7",
    pages = "751--754",
    year = "2022"
}

@article{Li_2012,
   title= "{Isospin breaking, coupled-channel effects, and $X(3872)$}",
   volume={86},
   ISSN={1550-2368},
   url={http://dx.doi.org/10.1103/PhysRevD.86.074022},
   DOI={10.1103/physrevd.86.074022},
   number={7},
   journal={Physical Review D},
   publisher={American Physical Society (APS)},
   author={Li, Ning and Zhu, Shi-Lin},
   year={2012},
   month=oct
}

@article{Yang:2011wz,
    author = "Yang, Zhong-Cheng and Sun, Zhi-Feng and He, Jun and Liu, Xiang and Zhu, Shi-Lin",
    title = "{The possible hidden-charm molecular baryons composed of anti-charmed meson and charmed baryon}",
    eprint = "1105.2901",
    archivePrefix = "arXiv",
    primaryClass = "hep-ph",
    doi = "10.1088/1674-1137/36/1/002",
    journal = "Chin. Phys. C",
    volume = "36",
    pages = "6--13",
    year = "2012"
}

@article{Wu:2010jy,
    author = "Wu, Jia-Jun and Molina, R. and Oset, E. and Zou, B. S.",
    title = "{Prediction of narrow $N^*$ and $\Lambda^*$ resonances with hidden charm above 4 GeV}",
    eprint = "1007.0573",
    archivePrefix = "arXiv",
    primaryClass = "nucl-th",
    doi = "10.1103/PhysRevLett.105.232001",
    journal = "Phys. Rev. Lett.",
    volume = "105",
    pages = "232001",
    year = "2010"
}

@article{Karliner:2015ina,
    author = "Karliner, Marek and Rosner, Jonathan L.",
    title = "{New Exotic Meson and Baryon Resonances from Doubly-Heavy Hadronic Molecules}",
    eprint = "1506.06386",
    archivePrefix = "arXiv",
    primaryClass = "hep-ph",
    reportNumber = "EFI-15-20, TAUP-2997-15",
    doi = "10.1103/PhysRevLett.115.122001",
    journal = "Phys. Rev. Lett.",
    volume = "115",
    number = "12",
    pages = "122001",
    year = "2015"
}

@article{Wang:2011rga,
    author = "Wang, W. L. and Huang, F. and Zhang, Z. Y. and Zou, B. S.",
    title = "{$\Sigma_c \bar{D}$ and $\Lambda_c \bar{D}$ states in a chiral quark model}",
    eprint = "1101.0453",
    archivePrefix = "arXiv",
    primaryClass = "nucl-th",
    doi = "10.1103/PhysRevC.84.015203",
    journal = "Phys. Rev. C",
    volume = "84",
    pages = "015203",
    year = "2011"
}

@article{Li:2014gra,
    author = "Li, Xue-Qian and Liu, Xiang",
    title = "{A possible global group structure for exotic states}",
    eprint = "1409.3332",
    archivePrefix = "arXiv",
    primaryClass = "hep-ph",
    doi = "10.1140/epjc/s10052-014-3198-3",
    journal = "Eur. Phys. J. C",
    volume = "74",
    number = "12",
    pages = "3198",
    year = "2014"
}

@article{Wu:2010vk,
    author = "Wu, Jia-Jun and Molina, R. and Oset, E. and Zou, B. S.",
    title = "{Dynamically generated $N^{*}$ and $\Lambda^*$ resonances in the hidden charm sector around 4.3 GeV}",
    eprint = "1011.2399",
    archivePrefix = "arXiv",
    primaryClass = "nucl-th",
    doi = "10.1103/PhysRevC.84.015202",
    journal = "Phys. Rev. C",
    volume = "84",
    pages = "015202",
    year = "2011"
}

@article{Chen_2015,
   title="{Identifying Exotic Hidden-Charm Pentaquarks}",
   volume={115},
   ISSN={1079-7114},
   url={http://dx.doi.org/10.1103/PhysRevLett.115.132002},
   DOI={10.1103/physrevlett.115.132002},
   number={13},
   journal={Physical Review Letters},
   publisher={American Physical Society (APS)},
   author={Chen, Rui and Liu, Xiang and Li, Xue-Qian and Zhu, Shi-Lin},
   year={2015},
   month=sep
}

@article{Chen_2017a,
   title="{Possible strange hidden-charm pentaquarks from and interactions}",
   volume={41},
   ISSN={1674-1137},
   url={http://dx.doi.org/10.1088/1674-1137/41/10/103105},
   DOI={10.1088/1674-1137/41/10/103105},
   number={10},
   journal={Chinese Physics C},
   publisher={IOP Publishing},
   author={Chen, Rui and He, Jun and Liu, Xiang},
   year={2017},
   month=sep, pages={103105}
}

@article{Chen_2019,
   title="{Strong LHCb evidence supporting the existence of the hidden-charm molecular pentaquarks}",
   volume={100},
   ISSN={2470-0029},
   url={http://dx.doi.org/10.1103/PhysRevD.100.011502},
   DOI={10.1103/physrevd.100.011502},
   number={1},
   journal={Physical Review D},
   publisher={American Physical Society (APS)},
   author={Chen, Rui and Sun, Zhi-Feng and Liu, Xiang and Zhu, Shi-Lin},
   year={2019},
   month=jul
}

@article{Chen_2016_2,
   title="{QCD sum rule study of hidden-charm pentaquarks}",
   volume={76},
   ISSN={1434-6052},
   url={http://dx.doi.org/10.1140/epjc/s10052-016-4438-5},
   DOI={10.1140/epjc/s10052-016-4438-5},
   number={10},
   journal={The European Physical Journal C},
   publisher={Springer Science and Business Media LLC},
   author={Chen, Hua-Xing and Cui, Er-Liang and Chen, Wei and Liu, Xiang and Steele, T. G. and Zhu, Shi-Lin},
   year={2016},
   month=oct
}

@article{Yang_2020,
   title="{Doubly charmed pentaquarks}",
   volume={101},
   ISSN={2470-0029},
   url={http://dx.doi.org/10.1103/PhysRevD.101.074030},
   DOI={10.1103/physrevd.101.074030},
   number={7},
   journal={Physical Review D},
   publisher={American Physical Society (APS)},
   author={Yang, Gang and Ping, Jialun and Segovia, Jorge},
   year={2020},
   month=apr
}

@article{ChenKan_2021,
   title="{Exploration of the doubly charmed molecular pentaquarks}",
   volume={103},
   ISSN={2470-0029},
   url={http://dx.doi.org/10.1103/PhysRevD.103.116017},
   DOI={10.1103/physrevd.103.116017},
   number={11},
   journal={Physical Review D},
   publisher={American Physical Society (APS)},
   author={Chen, Kan and Wang, Bo and Zhu, Shi-Lin},
   year={2021},
   month=jun
}

@article{ChenRui_2021,
   title="{Doubly charmed molecular pentaquarks}",
   volume={822},
   ISSN={0370-2693},
   url={http://dx.doi.org/10.1016/j.physletb.2021.136693},
   DOI={10.1016/j.physletb.2021.136693},
   journal={Physics Letters B},
   publisher={Elsevier BV},
   author={Chen, Rui and Li, Ning and Sun, Zhi-Feng and Liu, Xiang and Zhu, Shi-Lin},
   year={2021},
   month=nov, pages={136693}
}

@article{Duan_2024,
   title="{Doubly charmed pentaquark states in QCD sum rules}",
   volume={109},
   ISSN={2470-0029},
   url={http://dx.doi.org/10.1103/PhysRevD.109.094018},
   DOI={10.1103/physrevd.109.094018},
   number={9},
   journal={Physical Review D},
   publisher={American Physical Society (APS)},
   author={Duan, Feng-Bo and Wang, Qi-Nan and Yang, Zi-Yan and Chen, Xu-Liang and Chen, Wei},
   year={2024},
   month=may
}

@article{Lyu:2021qsh,
    author = "Lyu, Yan and Tong, Hui and Sugiura, Takuya and Aoki, Sinya and Doi, Takumi and Hatsuda, Tetsuo and Meng, Jie and Miyamoto, Takaya",
    title = "{Dibaryon with Highest Charm Number near Unitarity from Lattice QCD}",
    eprint = "2102.00181",
    archivePrefix = "arXiv",
    primaryClass = "hep-lat",
    reportNumber = "RIKEN-iTHEMS-Report-21,RIKEN-QHP-491",
    doi = "10.1103/PhysRevLett.127.072003",
    journal = "Phys. Rev. Lett.",
    volume = "127",
    number = "7",
    pages = "072003",
    year = "2021"
}

@article{Liu:2021pdu,
    author = "Liu, Ming-Zhu and Geng, Li-Sheng",
    title = "{Prediction of an $\Omega_{bbb} \Omega_{bbb}$ Dibaryon in the Extended One-Boson Exchange Model}",
    eprint = "2107.04957",
    archivePrefix = "arXiv",
    primaryClass = "hep-ph",
    doi = "10.1088/0256-307X/38/10/101201",
    journal = "Chin. Phys. Lett.",
    volume = "38",
    number = "10",
    pages = "101201",
    year = "2021"
}

@article{Mathur:2022ovu,
    author = "Mathur, Nilmani and Padmanath, M. and Chakraborty, Debsubhra",
    title = "{Strongly Bound Dibaryon with Maximal Beauty Flavor from Lattice QCD}",
    eprint = "2205.02862",
    archivePrefix = "arXiv",
    primaryClass = "hep-lat",
    reportNumber = "TIFR/TH/22-21, MITP-22-033",
    doi = "10.1103/PhysRevLett.130.111901",
    journal = "Phys. Rev. Lett.",
    volume = "130",
    number = "11",
    pages = "111901",
    year = "2023"
}

@article{Chen:2024xlw,
    author = "Chen, Ping and Liu, Zhan-Wei and Zhang, Zi-Le and Luo, Si-Qiang and Wang, Fu-Lai and Wang, Jun-Zhang and Liu, Xiang",
    title = "{Role of electromagnetic interactions in the $X(3872)$ and its analogs}",
    eprint = "2401.05989",
    archivePrefix = "arXiv",
    primaryClass = "hep-ph",
    doi = "10.1103/PhysRevD.109.094002",
    journal = "Phys. Rev. D",
    volume = "109",
    number = "9",
    pages = "094002",
    year = "2024"
}

@article{Yan:1992gz,
    author = "Yan, Tung-Mow and Cheng, Hai-Yang and Cheung, Chi-Yee and Lin, Guey-Lin and Lin, Y. C. and Yu, Hoi-Lai",
    title = "{Heavy quark symmetry and chiral dynamics}",
    reportNumber = "CLNS-92-1138, IP-ASTP-03-92",
    doi = "10.1103/PhysRevD.46.1148",
    journal = "Phys. Rev. D",
    volume = "46",
    pages = "1148--1164",
    year = "1992",
    note = "[Erratum: Phys.Rev.D 55, 5851 (1997)]"
}

@article{Wise:1992hn,
    author = "Wise, Mark B.",
    title = "{Chiral perturbation theory for hadrons containing a heavy quark}",
    reportNumber = "CALT-68-1765",
    doi = "10.1103/PhysRevD.45.R2188",
    journal = "Phys. Rev. D",
    volume = "45",
    number = "7",
    pages = "R2188",
    year = "1992"
}

@article{Burdman:1992gh,
    author = "Burdman, Gustavo and Donoghue, John F.",
    title = "{Union of chiral and heavy quark symmetries}",
    reportNumber = "UMHEP-365",
    doi = "10.1016/0370-2693(92)90068-F",
    journal = "Phys. Lett. B",
    volume = "280",
    pages = "287--291",
    year = "1992"
}

@article{Casalbuoni:1996pg,
    author = "Casalbuoni, R. and Deandrea, A. and Di Bartolomeo, N. and Gatto, Raoul and Feruglio, F. and Nardulli, G.",
    title = "{Phenomenology of heavy meson chiral Lagrangians}",
    eprint = "hep-ph/9605342",
    archivePrefix = "arXiv",
    reportNumber = "UGVA-DPT-1996-05-928, BARI-TH-96-237",
    doi = "10.1016/S0370-1573(96)00027-0",
    journal = "Phys. Rept.",
    volume = "281",
    pages = "145--238",
    year = "1997"
}

@article{Falk:1992cx,
    author = "Falk, Adam F. and Luke, Michael E.",
    title = "{Strong decays of excited heavy mesons in chiral perturbation theory}",
    eprint = "hep-ph/9206241",
    archivePrefix = "arXiv",
    reportNumber = "SLAC-PUB-5812, UCSD-PTH-92-14",
    doi = "10.1016/0370-2693(92)90618-E",
    journal = "Phys. Lett. B",
    volume = "292",
    pages = "119--127",
    year = "1992"
}

@article{Liu:2011xc,
    author = "Liu, Yan-Rui and Oka, Makoto",
    title = "{$\Lambda_c N$ bound states revisited}",
    eprint = "1103.4624",
    archivePrefix = "arXiv",
    primaryClass = "hep-ph",
    doi = "10.1103/PhysRevD.85.014015",
    journal = "Phys. Rev. D",
    volume = "85",
    pages = "014015",
    year = "2012"
}

@article{Wang_2024,
   title="{Doubly-charm and doubly-bottom pentaquark molecular states via the QCD sum rules}",
   volume={39},
   ISSN={1793-656X},
   url={http://dx.doi.org/10.1142/S0217751X24500672},
   DOI={10.1142/s0217751x24500672},
   number={17n18},
   journal={International Journal of Modern Physics A},
   publisher={World Scientific Pub Co Pte Ltd},
   author={Wang, Xiu-Wu and Wang, Zhi-Gang},
   year={2024},
   month=jun }

@article{MENG20231,
title = "{Chiral perturbation theory for heavy hadrons and chiral effective field theory for heavy hadronic molecules}",
journal = {Physics Reports},
volume = {1019},
pages = {1-149},
year = {2023},
note = {Chiral perturbation theory for heavy hadrons and chiral effective field theory for heavy hadronic molecules},
issn = {0370-1573},
doi = {https://doi.org/10.1016/j.physrep.2023.04.003},
url = {https://www.sciencedirect.com/science/article/pii/S0370157323001679},
author = {Lu Meng and Bo Wang and Guang-Juan Wang and Shi-Lin Zhu}
}

@article{Xing_2021,
   title="{The study of doubly charmed pentaquark $cc{\bar{q}}qq$ with the SU(3) symmetry}",
   volume={81},
   ISSN={1434-6052},
   url={http://dx.doi.org/10.1140/epjc/s10052-021-09730-4},
   DOI={10.1140/epjc/s10052-021-09730-4},
   number={11},
   journal={The European Physical Journal C},
   publisher={Springer Science and Business Media LLC},
   author={Xing, Ye and Niu, Yuekun},
   year={2021},
   month=nov }

@article{Dong_2021,
   title="{A survey of heavy–heavy hadronic molecules}",
   volume={73},
   ISSN={1572-9494},
   url={http://dx.doi.org/10.1088/1572-9494/ac27a2},
   DOI={10.1088/1572-9494/ac27a2},
   number={12},
   journal={Communications in Theoretical Physics},
   publisher={IOP Publishing},
   author={Dong, Xiang-Kun and Guo, Feng-Kun and Zou, Bing-Song},
   year={2021},
   month=oct, pages={125201} }

@misc{chen2025,
    title="{Doubly charmed pentaquark states with strangeness $S=0, -1$}", 
    author={Wei Chen and Feng-Bo Duan and Zi-Yan Yang and Qi-Nan Wang and Xu-Liang Chen and Qian Wang},
    year={2025},
    eprint={2509.01965},
    archivePrefix={arXiv},
    primaryClass={hep-ph},
    url={https://arxiv.org/abs/2509.01965}, 
}

@article{Li_2025,
   title="{Doubly-charmed pentaquark states in a mass splitting model}",
   volume={140},
   ISSN={2190-5444},
   url={http://dx.doi.org/10.1140/epjp/s13360-025-06732-z},
   DOI={10.1140/epjp/s13360-025-06732-z},
   number={8},
   journal={The European Physical Journal Plus},
   publisher={Springer Science and Business Media LLC},
   author={Li, Shi-Yuan and Liu, Yan-Rui and Shu, Cheng-Rui and Si, Zong-Guo},
   year={2025},
   month=aug
}

@article{Chen_2017b,
   title="{Heavy molecules and one-$\sigma/\omega$-exchange model}",
   volume={96},
   ISSN={2470-0029},
   url={http://dx.doi.org/10.1103/PhysRevD.96.116012},
   DOI={10.1103/physrevd.96.116012},
   number={11},
   journal={Physical Review D},
   publisher={American Physical Society (APS)},
   author={Chen, Rui and Hosaka, Atsushi and Liu, Xiang},
   year={2017},
   month=dec
}

@misc{liu2023,
      title="{Investigation of the analog of the $P_{c}$ states-the doubly charmed molecular pentaquarks}", 
      author={Xuejie Liu and Yue Tan and Xiaoyun Chen and Dianyong Chen and Hongxia Huang and Jialun Ping},
      year={2023},
      eprint={2312.04390},
      archivePrefix={arXiv},
      primaryClass={hep-ph},
      url={https://arxiv.org/abs/2312.04390}, 
}

@article{Shen_2023,
   title="{$P_{cc}^N$ states in a unitarized coupled-channel approach}",
   volume={83},
   ISSN={1434-6052},
   url={http://dx.doi.org/10.1140/epjc/s10052-023-11177-8},
   DOI={10.1140/epjc/s10052-023-11177-8},
   number={1},
   journal={The European Physical Journal C},
   publisher={Springer Science and Business Media LLC},
   author={Shen, Chao-Wei and Lin, Yong-hui and Meißner, Ulf-G.},
   year={2023},
   month=jan }

@article{Shimizu_2017,
   title="{Hidden charm pentaquark $P_{c}(4380)$ and doubly charmed baryon $\Xi_{cc}^{*}(4380)$ as hadronic molecule states}",
   volume={96},
   ISSN={2470-0029},
   url={http://dx.doi.org/10.1103/PhysRevD.96.094012},
   DOI={10.1103/physrevd.96.094012},
   number={9},
   journal={Physical Review D},
   publisher={American Physical Society (APS)},
   author={Shimizu, Yuki and Harada, Masayasu},
   year={2017},
   month=nov }

@article{Guo:2017vcf,
    author = "Guo, Zhi-Hui",
    title = "{Prediction of exotic doubly charmed baryons within chiral effective field theory}",
    eprint = "1708.04145",
    archivePrefix = "arXiv",
    primaryClass = "hep-ph",
    doi = "10.1103/PhysRevD.96.074004",
    journal = "Phys. Rev. D",
    volume = "96",
    number = "7",
    pages = "074004",
    year = "2017"
}

@article{Zhou:2018bkn,
    author = "Zhou, Qin-Song and Chen, Kan and Liu, Xiang and Liu, Yan-Rui and Zhu, Shi-Lin",
    title = "{Surveying exotic pentaquarks with the typical $QQqq\bar{q}$ configuration}",
    eprint = "1801.04557",
    archivePrefix = "arXiv",
    primaryClass = "hep-ph",
    reportNumber = "LZU-2018-01",
    doi = "10.1103/PhysRevC.98.045204",
    journal = "Phys. Rev. C",
    volume = "98",
    number = "4",
    pages = "045204",
    year = "2018"
}

@article{Wang:2018lhz,
    author = "Wang, Zhi-Gang",
    title = "{Analysis of the doubly heavy baryon states and pentaquark states with QCD sum rules}",
    eprint = "1808.09820",
    archivePrefix = "arXiv",
    primaryClass = "hep-ph",
    doi = "10.1140/epjc/s10052-018-6300-4",
    journal = "Eur. Phys. J. C",
    volume = "78",
    number = "10",
    pages = "826",
    year = "2018"
}

@article{Goitein:1967kaq,
    author = "Goitein, M. and Dunning, Jr., J. R. and Wilson, Richard",
    title = "{Comparison of Elastic Electron-Proton Scattering Cross Sections with Some Theoretical Predictions}",
    doi = "10.1103/PhysRevLett.18.1018",
    journal = "Phys. Rev. Lett.",
    volume = "18",
    number = "23",
    pages = "1018",
    year = "1967"
}

@article{AUERBACH:1972yql,
    author = "AUERBACH, NAFTALI and HUFNER, JORG and KERMAN, A. K. and SHAKIN, C. M.",
    title = "{A Theory of Isobaric Analog Resonances}",
    doi = "10.1103/RevModPhys.44.48",
    journal = "Rev. Mod. Phys.",
    volume = "44",
    pages = "48--125",
    year = "1972"
}

@misc{An:2025qfw,
    author = "An, Hong-Tao and Li, Yu-Shuai",
    title = "{Systematic investigation of the spectroscopy and decay behaviors of doubly-charmed pentaquarks}",
    eprint = "2512.08643",
    archivePrefix = "arXiv",
    primaryClass = "hep-ph",
    month = "12",
    year = "2025"
}

\end{document}